\documentclass[a4paper,fleqn,usenatbib]{mnras}

\usepackage[T1]{fontenc}
\usepackage{ae,aecompl}

\usepackage[pdftex]{graphicx}	
\usepackage{here}
\usepackage{amsmath}	
\usepackage{amssymb}	

\usepackage{lscape}
\usepackage{bm}
\usepackage{comment}
\usepackage{newtxtext,newtxmath}

\title[AD effect on the first star formation]{Non-ideal magnetohydrodynamic simulations of the first star formation: the effect of ambipolar diffusion}

\author[K. E. Sadanari et al.]{
Kenji Eric Sadanari,$^{1}$\thanks{E-mail: k.sadanari@astr.tohoku.ac.jp}
Kazuyuki Omukai,$^{1}$
Kazuyuki Sugimura,$^{2,3}$
Tomoaki Matsumoto,$^{4}$
 and Kengo Tomida$^{1}$
\\
$^{1}$Astronomical Institute, Tohoku University, Aoba, Sendai, Miyagi 980-8578, Japan\\
$^{2}$The Hakubi Center for Advanced Research, Kyoto University, Yoshida-honmachi, Sakyo-ku, Kyoto 606-8501, Japan\\
$^{3}$Department of Physics, Graduate School of Science, Kyoto University, Sakyo, Kyoto 606-8502, Japan\\
$^{4}$Faculty of Sustainability Studies, Hosei University, Fujimi, Chiyoda, Tokyo 102-8160, Japan\\
}

\date{Accepted XXX. Received YYY; in original form ZZZ}

\pubyear{2022}

\begin{document}
\label{firstpage}
\pagerange{\pageref{firstpage}--\pageref{lastpage}}
\maketitle

\begin{abstract}
In the present-day universe, magnetic fields play such essential roles in star formation as angular momentum transport and outflow driving, which control circumstellar disc formation/fragmentation and also the star formation efficiency. 
While only a much weaker field has been believed to exist in the early universe, recent theoretical studies find that strong fields can be generated by turbulent dynamo during the gravitational collapse.
Here, we investigate the gravitational collapse of a cloud core ($\sim 10^{3}\ \rm cm^{-3}$) up to protostar formation ($\sim 10^{20}\ \rm cm^{-3}$) by non-ideal magnetohydrodynamics (MHD) simulations considering ambipolar diffusion (AD), the dominant non-ideal effects in the primordial-gas.
We systematically study rotating cloud cores either with or without turbulence and permeated with uniform fields of different strengths. 
We find that AD can slightly suppress the field growth by dynamo especially on scales smaller than the Jeans-scale at the density range $10^{10}-10^{14}\ \rm cm^{-3}$, 
while we could not see the AD effect on the temperature evolution, since the AD heating rate is always smaller than compression heating.  
The inefficiency of AD makes the field as strong as $10^{3}-10^{5} \rm\ G$ near the formed protostar, much stronger than in the present-day cases, even in cases with initially weak fields. 
The magnetic field affects the inflow motion when amplified to the equipartition level with turbulence on the Jeans-scale, although disturbed fields do not launch winds.
This might suggest that dynamo amplified fields have smaller impact on the dynamics in the later accretion phase than other processes such as ionisation feedback.
\end{abstract}

\begin{keywords}
stars:formation, stars: Population$\rm I\hspace{-.1em}I\hspace{-.1em}I$, stars:magnetic field
\end{keywords}


\section{Introduction}\label{Sec:introduction}
Formation of first stars marks a fundamental turning point in the history of the universe.
The light emitted by them ends the cosmic dark ages (e.g., \citealp{Barkana_Loeb2001}).
In particular, strong UV radiation from massive first stars heats the intergalactic medium (IGM)/ interstellar medium (ISM), and affects subsequent star-formation 
(e.g., \citealp{Bromm2001}; \citealp{Ciardi_Ferrara2005}; \citealp{Bromm2011}).
Their supernova (SN) explosions also enrich the primordial gas with the first heavy elements, thereby triggering the transition from Pop III to Pop II star formation 
(e.g., \citealp{Heger2003}; \citealp{Umeda_Nommoto2003}; \citealp{Greif2010}).
Massive and close binary systems, if formed among this population, may evolve to binary black holes (BHs) of a few $10\ M_{\odot}$, whose merger events are recently observed by gravitational waves
(e.g., \citealp{Kinugawa2014}, \citeyear{Kinugawa2016}; \citealp{Hartwig2016}; \citealp{Abbott2016}). 
                           
First stars form at redshift $z \sim 20$-$30$ in small dark matter (DM) halos known as minihalos of $10^{5}$-$10^{6}\ M_{\odot}$ 
(\citealp{Couchman_Rees1986}; \citealp{Yoshida2003}; \citealp{Greif2015}).
Inside a minihalo, a massive gas core of mass $\sim 10^{3}\ M_{\odot}$ collapses at a temperature of few hundred $\rm K$ due to the H$_2$ cooling to form a protostar at the center (\citealp{Abel2002}; \citealp{Bromm2002}; \citealp{Yoshida2008}).
After the formation, the protostar grows in mass by accretion of surrounding gas
at a high rate $\sim 10^{-3}\ M_{\odot}\rm yr^{-1}$ reflecting the high gas temperature (\citealp{Stahler_Palla1986}; \citealp{Omukai_Nishi1998}).
The mass of the forming star is set when the accretion is terminated, usually by the stellar radiative feedback, and reaches as massive as a few $10$ - a few $100\ \rm M_{\odot}$
(\citealp{Omukai_Palla2003}; \citealp{McKee_Tan2008}; \citealp{Hosokawa2016}; \citealp{Stacy2016}).
Recent simulations also show that
the first stars generally form as binary or multiple protostellar systems due to fragmentation of circumstellar discs (e.g., \citealp{Smith2011}; \citealp{Greif2012}; \citealp{Stacy_Bromm2013}; \citealp{Susa2019}; \citealp{Chon_Hosokawa2019}; \citealp{Sugimura2020}; \citealp{Kimura2020}).
Turbulence, if presents, further promotes the disc fragmentation and causes a wider mass distribution of a few $10^{-3}\ M_{\odot}$ to a few $10\ \rm M_{\odot}$
(\citealp{Wollenberg2020}).

The presence of magnetic fields can significantly alter those conclusions on the nature of first stars.
In present-day star forming regions, strong coherent magnetic fields of several $\mu \rm G$, which is comparable to the gravitational energy on the core scale, are observed (e.g., \citealp{Heiles_Troland2005}; \citealp{Troland_Crutcher2008}). 
Such a strong field effectively transports the angular momentum by magnetic braking, i.e., braking of the gas motion by magnetic tension (\citealp{Mouschovias_Paleologou1979}), 
and significantly reduces the circumstellar disc size, thereby suppressing its fragmentation.
Magnetic fields can also drive outflows in various ways, such as magnetocentrifugal winds (\citealp{Blandford_Payne1982}) and magnetic pressure winds 
(\citealp{Tomisaka2002}; \citealp{Banerjee_Pudritz2006}; \citealp{Machida2008a}).
 By ejecting the part of the accreting gas,
those MHD winds transport the angular momentum outward and also reduces the mass of formed stars (e.g., \citealp{Machida_Hosokawa2013}). 
According to numerical studies, the strength of those effects depends on the field configuration in a way that the magnetic braking efficiency is reduced for turbulent disturbed field by the misalignment between the field and rotation axes (e.g., \citealp{Joos2013}) and reconnection diffusion (e.g., \citealp{Santos-Lima2013}).
\cite{Gerrard2019} also suggested that some coherent field is needed for driving outflows.

The nature of magnetic fields in the early universe are highly uncertain,
while theoretically at least weak seed magnetic fields are expected to exist.
A promising mechanism for their generation is the so-called Biermann battery mechanism (\citealp{Biermann1950}; \citealp{Biermann1951}),
which can generate seed fields of $10^{-20}$-$10^{-18}\ \rm G$ (in the physical unit).
This operates in the case where the electron-density and pressure gradients are not parallel, 
which can occur in various astrophysical situations such as supernovae explosion (\citealp{Hanayama2005}), galaxy formation (\citealp{Kulsrud1997}), reionisation (\citealp{Gnedin2000}; \citealp{Attia2021}), ionisation fronts around massive stars 
(\citealp{Langer2003}; \citealp{Doi_Susa2011}), virialization shock in a minihalo (\citealp{Xu2008}), and streaming of first cosmic rays in the universe (\citealp{Ohira2020}, \citeyear{Ohira2021}).
Seed fields can be created by other mechanisms: for example, $\sim 10^{-24}$ G field on a few $\rm Mpc$ scale is generated by second-order couplings between photons and electrons due to cosmological fluctuations before cosmological recombination at $z\sim 1100$ (\citealp{Saga2015}).
Although a weak field of order of $10^{-18}\ \rm G$ or less is not dynamically important, it can be subsequently amplified by dynamo mechanism driven, e.g., by turbulence.
The turbulent dynamo is expected to be the most effective on the smallest scales  (see Sec. \ref{sec_dynamo}) and can generate a field as strong as $\sim 10^{-6}\ \rm G$ (\citealp{Schleicher2010}; \citealp{Sur2010}; \citealp{Schober2012b}; \citealp{Xu_Lazarian2016}). 

Many authors have studied the first star formation by way of numerical magnetohydrodynamics (MHD) simulations 
(\citealp{Machida2008}; \citealp{Machida_Doi2013}; \citealp{Sharda2020b}, \citeyear{Sharda2020a}; \citealp{Stacy2022}; \citealp{Prole2022}; \citealp{Hirano2022}; \citealp{Saad2022}).
For example, \cite{Stacy2022} studied gravitational collapse of star-forming cloud cores up to the protostellar accretion phase starting from the cosmological initial conditions, assuming that a magnetic field is amplified by the small-scale dynamo at the initial stages.
They concluded that the amplified magnetic field effectively suppresses the disc fragmentation, thereby leading to more top-heavy initial mass function (IMF) than in the case without the field.
In most simulations of this kind, however, the ideal MHD is assumed \footnote{The Ohmic dissipation was included in 
simulations of \citealp{Machida_Doi2013}, but turned out to be unimportant.} and non-ideal MHD effects such as the Ohmic dissipation, ambipolar diffusion (AD), and Hall effect are not taken into account. 

In the present-day star formation, magnetic fields effectively dissipate by the Ohmic dissipation and AD.
In fact, when the field dissipation is taken into account, 
the discs around the protostars become larger in 3D simulations
as the magnetic braking becomes less effective
(\citealp{Machida2007}; \citealp{Tomida2013},\citeyear{Tomida2015}; \citealp{Tsukamoto2015}; \citealp{Masson2016}).
While the Hall effect does not participate in the field energy dissipation, it affects the angular momentum distribution and thus the disc structure (\citealp{Tsukamoto2017}).
In the first star formation, those non-ideal MHD effects are expected to be less effective due to higher ionisation degree in the primordial gas owing to the higher temperature and the absence of dust (\citealp{Maki_Susa2004}, \citeyear{Maki_Susa2007}).
Still, AD is expected to operate once the field becomes strong enough since its resistivity is proportional to the square of the field strength. 
Indeed, based on one-zone calculations, it is claimed that AD can affect the thermal and then indirectly dynamical evolution of a collapsing cloud once the field energy becomes comparable to the gravitational energy (\citealp{Schleicher2009}; \citealp{Sethi2010}; \citealp{Nakauchi2019}).
AD, however, depends not only on the field strength but also on its structure, which cannot be properly taken into account in the one-zone calculations.

Here, to clarify how the magnetic field is amplified and affects the gas dynamics considering the AD effect,
we perform, for the first time, 3D non-ideal MHD simulations by taking into account of AD for the collapsing primordial gas clouds.
The paper is organised as follows.
In Section \ref{sec_dynamo}, we briefly introduce how the small-scale dynamo amplifies a weak seed field.
In Section \ref{sec_method}, we describe the numerical method and initial set-up of the calculations.
We present simulation results in Section \ref{sec_result}: the evolution of a collapsing cloud with pure rotation (Section \ref{sec_rot}), the cases with both rotation and turbulence (Section \ref{sec_turb}), 
the extent of the AD effect on its evolution (Section \ref{sec_AD}), and the resolution dependence of the results (Section \ref{Appendix_A}).
In Section \ref{Sec:discussion_conclusion}, we summarize our findings and discuss the possible influence of amplified magnetic fields on the first star formation.
\section{magnetic field amplification by small-scale dynamo}\label{sec_dynamo}
As mentioned in Sec. \ref{Sec:introduction}, weak seed magnetic fields of $\sim 10^{-20}$-$10^{-18}\ \rm G$ in the early universe can be amplified by stretching and folding turbulent motions.
The amplification is particularly efficient on smaller scales, where the eddy turnover time of turbulence is shorter, so is called the small-scale dynamo.
According to theories, this works efficiently in a collapsing primordial-gas cloud
(\citealp{Batchelor1950}; \citealp{Kazantsev1968}; \citealp{Kulsrud1992}; \citealp{Schekochihin2004}; \citealp{Schleicher2010}; \citealp{Schober2012b}; \citealp{Sur2010}; \citealp{Turk2012}; \citealp{Xu_Lazarian2016}).
The small-scale dynamo proceeds in two stages,
the kinematic stage where the magnetic field at a small scale is amplified exponentially without being hindered by the back-reaction of magnetic forces, and the subsequent non-linear stage where the magnetic field on larger scale is amplified only gradually due to the field back-reaction.
In this section, we briefly review how the small-scale dynamo works in these two stages in the context of 
the field amplification in the current numerical simulation. 
For detailed arguments, see \cite{Xu_Lazarian2016} and \cite{McKee2020}.

In the kinematic stage, being much smaller than turbulence in energy, 
the magnetic field is amplified by turbulence on the scale $l$ in the eddy turnover time $t_{\rm eddy}(l) \sim l/V_{\rm turb}(l)$, where $V_{\rm turb}(l)$ is the turbulent speed at the scale $l$.
For the Kolmogorov turbulence with the scale dependence of
\begin{equation} \label{eq_Kol}
V_{\rm turb}(l) \propto l ^{1/3},
\end{equation}
the amplification rate is given by $\Gamma(l)\sim t^{-1}_{\rm eddy} \propto l^{-2/3}$.
This suggests that the magnetic field growth is the fastest on the smallest scale of turbulence
set by the larger of the viscous scale $l_{\rm \nu}$ and the resistivity scale $l_{\rm \eta}$.
In the case of primordial gas, $l_{\rm \eta}$ is determined by AD and 
considerably smaller than $l_{\rm \nu}$,
since the field in the kinematic stage is still small.
Therefore, the specific magnetic energy density $\mathcal{E}_{\rm mag}$ exponentially grows as 
\begin{equation} \label{eq_kine}
\mathcal{E}_{\rm mag}(t) \propto \exp{(C\Gamma(l_{\rm \nu})t)} = \exp{\left( C \frac{t}{t_{\rm eddy}(l_{\rm \nu})}\right)},
\end{equation}
where the constant $C$ represents the efficiency of amplification (e.g., \citealp{Xu_Lazarian2016}; \citealp{McKee2020}; \citealp{Stacy2022}), with $C=37/36\sim 1$ in the case of ideal MHD ($l_{\rm \nu}/l_{\rm \eta}\gg 1$; \citealp{Schober2012a}).
Since the eddy turnover time on the viscous scale $t_{\rm eddy}(l_{\rm \nu})$
is sufficiently shorter than the free-fall time, 
the amplification by the kinematic dynamo is much more efficient than the amplification by the global compression in a collapsing cloud.

Note, however, that lower field growth rates are observed in numerical simulations because of the limited resolution.
Since the field dissipates numerically at the cell scale (e.g., \citealp{Lesaffre_Balbus2007}), rather than physically at much smaller viscosity/resistivity scales, the field is amplified only with a longer eddy turnover time $t_{\rm eddy}$, or correspondingly with a much lower efficiency $C\ll 1$ (\citealp{Sur2010}; \citealp{Federrath2011a}; \citealp{Stacy2022}) in eq. (\ref{eq_kine}).
Previous studies have shown that the Jeans length must be resolved with more than $32$-$64$ cells
to capture the kinematic dynamo (although with a lower efficiency) in the collapsing cloud (\citealp{Sur2010}; \citealp{Federrath2011b}; \citealp{Turk2012}).

Once the magnetic field becomes comparable to the turbulence in energy on the smallest scale,
the dynamo enters the non-linear stage.
Afterwards, the back-reaction of the field prevents its exponential growth on the small scales where the magnetic and turbulent energies are in equipartition.
On larger scales, however, the field is still much weaker than the turbulence in energy, 
and can continues to exponentially grow without back-reaction. 
As a result, the peak scale of the field energy $l_{\rm p}$ shifts towards a larger scale, below which the equipartition has been reached.
Through this process, the field energy density grows linearly in time as (e.g., \citealp{Xu_Lazarian2016}, \citeyear{Xu_Lazarian2020})
\begin{equation} \label{eq_non_lin}
\mathcal{E}_{\rm mag}(t) \propto t .
\end{equation}
In the non-linear stage, the dynamo amplification is in general slower than the amplification by the global compression in the collapsing cloud, 
and thus the field amplification can be mainly attributed to the latter mechanism.

The dynamo in the non-linear stage comes to an end when the peak scale of the field energy reaches the driving scale of the turbulence.
Since the driving scale in a collapsing cloud is roughly the Jeans scale $L_{\rm J}$ (\citealp{Federrath2011b}),
the field strength at the (full-scale) equipartition $B_{\rm eq}$ can be estimated from $B_{\rm eq}^2/(8\pi \rho)=V_{\rm turb}^2/2\simeq V_{\rm turb}(L_{\rm J})^2/2$ as
\begin{equation} \label{eq_Beq}
B_{\rm eq} = \sqrt{4\pi\rho} V_{\rm turb}(L_{\rm J}).
\end{equation}
In numerical simulations with forced subsonic turbulence, the field saturates around $0.7 B_{\rm eq}$ (\citealp{Haugen2004}; \citealp{Federrath2011a}; \citealp{Brandenburg2014}). 
After the end of the non-linear stage, the field grows solely by the gravitational compression without the aid of dynamo action.

\section{Numerical Method}\label{sec_method}
\subsection{Code description}
We perform magnetohydrodynamics simulations with \textsc{SFUMATO-RT} (\citealp{Sugimura2020}), an extension of 
adaptive mesh refinement (AMR) code \textsc{SFUMATO} (\citealp{Matsumoto2007}; \citealp{Matsumoto2015}).
In the code, thermal and chemical evolution of the primordial gas is solved consistently with the gas dynamics.
The numerical method and basic equations are the same as in \cite{Sadanari2021} except that we take into account AD of magnetic fields in this study.

The governing equations are as follows: 
 the mass conservation, 
\begin{equation} \label{eq1}
\frac{\partial \rho}{\partial t} + \nabla  \cdot \left( \rho \bm{v} \right)=0,
\end{equation}
the equation of motion, 
\begin{equation} \label{eq2}
\rho \frac{\partial \bm{v}}{\partial t} + \rho \left( \bm{v} \cdot \nabla \right) \bm{v} = - \nabla p- \frac{1}{4 \pi} \bm{B} \times \left( \nabla \times \bm{B}\right) - \rho \nabla \phi,
\end{equation}
the gas energy equation, 
\begin{multline} \label{eq3}
\frac{\partial e}{\partial t} + \nabla \cdot \left[  \left(  e + p + \frac{|\bm{B}|^{2}}{8\pi}   \right) \bm{v} -  \frac{1}{4\pi} \bm{B} \left(  \bm{v} \cdot \bm{B}\right)  \right.\\
\left. - \frac{\eta_{\rm AD}}{4\pi | \bm{B}|^2} \left( \bm{B} \times \left(  \left(\nabla \times \bm{B} \right) \times \bm{B} \right) \right) \times \bm{B} \right]  
=  -\rho \bm{v} \cdot \nabla \phi -\Lambda,
\end{multline}
the induction equation including AD, 
\begin{equation} \label{eq4}
\frac{\partial \bm{B}}{\partial t} = \nabla \times \left(  \bm{v}\times \bm{B}    
- \frac{\eta_{\rm AD}}{|\bm{B}|^2} \bm{B} \times \left(  \left( \nabla \times \bm{B}\right) \times \bm{B}\right) \right),
\end{equation}
the solenoidal constraint, 
\begin{equation} \label{eq5}
\nabla \cdot \bm{B} = 0,
\end{equation}
and the Poisson equation for the gravity,
\begin{equation} \label{eq6}
\nabla ^{2} \phi = 4 \pi G \rho,
\end{equation}
where $\rho$, $p$, $\bm{v}$, $\bm{B}$, $\phi$, $e$,  $\eta_{\rm AD}$, $\Lambda$ are the gas density, gas pressure, gas velocity, magnetic field, gravitational potential, total gas energy per unit volume, resistivity of AD and net cooling rate per unit volume, respectively.
The energy density $e$ is given by
\begin{equation}
e = \frac{1}{2}\rho |\bm{v}|^{2} +\frac{p}{\gamma-1}+\frac{1}{8\pi}|\bm{B}|^{2}, 
\end{equation}
where $\gamma$ is the adiabatic index, which depends on the chemical composition and gas temperature (e.g., \citealp{Omukai_Nishi1998}).

For the AD term in the induction equation,
we take operator-splitting approach with the single-fluid approximation  
(\citealp{Mac1995}; \citealp{Duffin_Pudritz2008}; \citealp{Masson2012}; \citealp{Tomida2015}).
We have confirmed that numerically generated divergence errors i.e., $\Delta x |\nabla \cdot \bm{B}|/|\bm{B}|$, where $\Delta x $ is the cell size, remain small always below a few per cent, even in cases with the AD effect.
Other non-ideal MHD effects, i.e., the Ohmic dissipation and Hall effect, are not included as their effects are minor compared with AD (e.g., \citealp{Nakauchi2019}). 
We calculate the AD resistivity $\eta_{\rm AD}$ as in \cite{Nakauchi2019}:
\begin{equation} 
\label{eq_etaAD}
\eta_{\rm AD} = \frac{c^{2}}{4\pi}\frac{\sigma_{\rm P}}{\sigma^2_{\rm P}+\sigma^2_{\rm H}} - \frac{c^{2}}{4\pi \sigma_{\rm O}},
\end{equation}
where $\sigma_{\rm P},\ \sigma_{\rm H}$ and $\sigma_{\rm O}$ are Pedersen, Hall and Ohmic conductivities, respectively, and can be written as
\begin{equation} \label{}
\sigma_{\rm P} = \left( \frac{c}{B}\right)^2 \sum_{\rm \nu} \frac{\rho_{\rm \nu}\tau_{\rm \nu} \omega^2_{\rm \nu}}{1 + \tau^2_{\rm \nu} \omega^2_{\rm \nu}},
\end{equation}
\begin{equation} \label{}
\sigma_{\rm H} = \left( \frac{c}{B}\right)^2  \sum_{\rm \nu}  \frac{q_{\rm \nu}}{|q_{\rm \nu}|} \frac{\rho_{\rm \nu}\omega_{\rm \nu}}{1 + \tau^2_{\rm \nu} \omega^2_{\rm \nu}},
\end{equation}
and
\begin{equation} \label{}
\sigma_{\rm O} = \left( \frac{c}{B}\right)^2  \sum_{\rm \nu}  \rho_{\rm \nu} \tau_{\rm \nu} \omega^2_{\rm \nu},
\end{equation}
with subscript $\rm \nu$ representing a species of charged particle which has the charge $q_{\rm \nu}$ and mass $m_{\rm \nu}$.
For each charged species, $\rho_{\rm \nu}$ is the mass density,  $\omega_{\rm \nu}=e|q_{\rm \nu}|B/(m_{\rm \nu}c)$ the cyclotron frequency, and 
$\tau_{\rm \nu}$ the collision timescale between charged and neutral particles (\citealp{Nakano_Umebayashi1986}):
\begin{equation} \label{}
\tau^{-1}_{\rm \nu} =  \sum_{\rm n} \tau^{-1}_{\rm \nu,n} =\sum_{\rm n} \frac{\mu_{\rm \nu,n}n_{\rm \nu}n_{\rm n} \langle \sigma v\rangle_{\rm \nu, n}}{\rho_{\rm \nu}},
\end{equation}
where the subscript $\rm n$ represents a species of neutral particle, i.e., $\rm H$, $\rm H_{2}$, or $\rm He$, and 
$\mu_{\rm \nu,n}$ and $\langle \sigma v\rangle_{\rm \nu, n}$ are the reduced mass and collision rate coefficient between $\rm \nu$ and $\rm n$, respectively. 

The net cooling rate $\Lambda$ in the energy equation (eq.\ref{eq3}) consists of the line cooling rate $\Lambda_{\rm line}$ ($\rm H_{2}$ and $\rm HD$), continuum cooling rate $\Lambda_{\rm cont}$ ($\rm H$ free-bound emission, $\rm H^{-}$ free-bound emission, $\rm H^{-}$ free-free emission, $\rm H$ free-free emission, $\rm H_{2}$-$\rm H_{2}$ collision-induced emission, and $\rm H_{2}$-$\rm He$ collision-induced emission), and chemical cooling/heating rate $\Lambda_{\rm chem}$ ($\rm H$ ionisation/recombination and $\rm H_{2}$ dissociation/formation). For details, see \cite{Sadanari2021}.

For the chemical network of the primordial gas, we take into account 30 chemical reactions among 12 species:
$\rm H,\ H_{2},\ H^{+},\ e,\ H^{-},\ H_{2}^{+},\ H_{3}^{+},\ D,\ HD,\ D^{+}, Li,\ Li^{+}$.
We assume that all the helium is neutral with concentration $y(\rm{He}) = 9.77\times10^{-2}$.
We adopt the minimum chemical network presented in \cite{Nakauchi2019},
which can correctly reproduce the temperature in the primordial gas
as well as the ionisation degree, needed for the calculation of the AD resistivity.

We take the calculation box size $L_{\rm{box}}=4\times10^{6}\ \rm{au}$, four times larger than the cloud radius $R_{\rm cl}$ (see below). 
We initially set base grids with $N_{\rm{base}}=256$ cells in each direction.
The cell is refined when the cell size exceeds 1/64 of the local Jeans length so as to capture the dynamo action 
(\citealp{Sur2010}; \citealp{Federrath2011b}; \citealp{Turk2012}).
The maximum refinement level is $l_{\rm{max}}=27$, and thus the minimum size of the cell is 
$\Delta x_{\rm{min}}=L_{\rm{box}}/N_{\rm{base}}\times2^{-l_{\rm{max}}} \simeq 1.2\times10^{-4}\ \rm{au}$.
The simulation is terminated when the first protostar is formed in the box at $n_{\rm H} \sim 10^{20-21}\ \rm cm^{-3}$. 
\subsection{Initial conditions}\label{sec_init}
Cosmological simulations suggest that gas accumulates in the center of minihalos, forming dense cloud cores, called loitering cloud cores 
(\citealp{Bromm1999}). 
Here, we take a spherical cloud core that mimics the loitering core embedded in a homogeneous medium as the initial condition of our calculation, as in \cite{Sadanari2021}. 
we adopt the density profile enhanced 1.4 times from that of the critical Bonnor-Ebert sphere (\citealp{Ebert1955}; \citealp{Bonnor1956}), i.e., a hydrostatic equilibrium configuration with external pressure on the verge of gravitational collapse.
The initial cloud has the central number density $n_{\rm c,0}=1.4\times10^{3}\ \rm{cm^{-3}}$ with the uniform temperature $T_{\rm{init}}=198\ \rm{K}$,
which is the temperature of a collapsing cloud core when the density reaches $n_{\rm c,0}$
obtained from a one-zone calculation.
The initial radius and mass are $R_{\rm cl}=1.1\times10^{6}\ \rm au$ and $M_{\rm{cl}}=5.5\times10^{3}\  M_{\odot}$, respectively.
As the boundary condition, the ambient uniform gas outside the initial cloud radius $R_{\rm cl}$ is fixed to the initial value.
We add a small (one percent of) $m=2$-mode density perturbation to it as in \cite{Sadanari2021} to break the symmetry.
We summarise initial properties of the simulated cloud core in Table \ref{table_cloud}.

We consider two kinds of the initial velocity field inside the core, (i) that with rigid rotation only ({\it pure rotation} cases) and (ii) that with turbulent motions in addition to the rigid rotation ({\it turbulent} cases).
Rotation with energy $E_{\rm rot}/|E_{\rm g}|=10^{-2}$ is assumed in the all cases, following the cosmological simulations of e.g., \citealp{Hirano2014}, 
which suggested that first-star forming clouds rotate at roughly $E_{\rm rot}/|E_{\rm g}|=10^{-2}-10^{-1}$.
In the turbulent cases, we also add
a turbulent velocity field with power spectrum $P(k) \propto k^{-4}$, where $k$ is the wavenumber, consistent with the Larson's scaling relation (\citealp{Larson1981}).
According to the cosmological simulations (\citealp{Greif2012}; \citealp{Stacy_Bromm2013}; \citealp{Stacy2022}), 
the strength of turbulence in a first-star forming cloud core ($n_{\rm c } \sim 10^{3}\ \rm cm^{-3}$) is around the average Mach number of $0.6-0.8$.
Here, the turbulent energy inside the core is set at 3 $\%$ of the gravitational energy, i.e., $E_{\rm turb}/|E_{\rm g}| = 3\times10^{-2}$, corresponding to an average Mach number $\mathcal{M} = 0.4$.
Since the turbulence can be amplified to $\mathcal{M} \sim 1$ by the gravitational compression (\citealp{Higashi2021}, \citeyear{Higashi2022}), 
the difference in the initial turbulent strength does not significantly alter the simulation results.

Furthermore, we put a uniform magnetic field $B_{\rm init}$ parallel to the rotation axis, 
and consider cases with six different field energies, 
$E_{\rm mag}/|E_{\rm g}|= 0,$ $2\times 10^{-9}\ (B_{\rm init}=10^{-9}\ \rm G,  {\rm respectively}),$ $2\times10^{-7}\ (10^{-8}\ \rm G),$ $2\times10^{-5}\ (10^{-7}\ \rm G),$ $2\times10^{-3}\ (10^{-6}\ \rm G),$ and 
$ 2\times10^{-1}\ (10^{-5}\ \rm G)$ both for the pure rotation and turbulent cases.
Table \ref{table_model} summarises the 12 runs examined in this study.
\begin{table}
 \caption{Initial properties of the simulated cloud core}
 \label{table_cloud}
 \centering
  \begin{tabular}{lc}
   \hline \hline
   Parameter  &  Value\\
 
   \hline 
   mass $M_{\rm cl}$               & $5.5\times10^{3}\ M_{\odot}$\\
   radius $R_{\rm cl}$             &  $1.1\times10^{6}\ \rm au$\\
   central number density $n_{\rm c,0}$  & $1.4\times10^{3}\ \rm cm^{-3}$\\
   temperature $T_{\rm init}$       & $198\ \rm K$\\
   ratio of thermal to gravitational energies $E_{\rm th}/|E_{\rm g}|$ & $0.6$\\
   \hline
  \end{tabular}
\end{table}
\begin{table}
 \caption{Model parameters.}
 \label{table_model}
 \centering
  \begin{tabular}{lccccc}
   \hline \hline
   Model  &  $ E_{\rm{rot}}/|E_{\rm g}|$  &  $E_{\rm{turb}}/|E_{\rm g}|$  &  $E_{\rm{mag}}/|E_{\rm g}|$ & $B_{\rm init}\ \rm[G]$ & $\mu_{0} $\\
 
   \hline 
   \multicolumn{6}{l}{pure rotation cases}\\
   \hline
   T0M0   ..... & $10^{-2}$  & $0                      $   &  $0                      $       & $ 0          $    &   $\infty$ \\
   T0M9   ..... & $10^{-2}$  & $0                      $   &  $2\times 10^{-9}$       &  $10^{-9}$     &  27000 \\
   T0M7   ..... & $10^{-2}$  & $0                      $   &  $2\times 10^{-7}$       &  $10^{-8}$     &  2700  \\
   T0M5   ..... & $10^{-2}$  & $0                      $   &  $2\times 10^{-5}$       &  $10^{-7}$     &  270  \\
   T0M3   ..... & $10^{-2}$ & $0                      $   &  $2\times 10^{-3}$       &  $10^{-6}$     &  27    \\
   T0M1   ..... & $10^{-2}$ & $0                      $   &  $2\times 10^{-1}$       &  $10^{-5}$     &  2.7\\
   \hline 
   \multicolumn{6}{l}{turbulent cases}\\
   \hline
   T2M0   ..... & $10^{-2}$  & $3\times 10^{-2}$   &  $0                      $       & $ 0          $    &   $\infty$ \\
   T2M9   ..... & $10^{-2}$  & $3\times 10^{-2}$   &  $2\times 10^{-9}$       &  $10^{-9}$     &  27000 \\
   T2M7   ..... & $10^{-2}$  & $3\times 10^{-2}$   &  $2\times 10^{-7}$       &  $10^{-8}$     &  2700  \\
   T2M5 ..... & $10^{-2}$  & $3\times 10^{-2}$   &  $2\times 10^{-5}$       &  $10^{-7}$     &  270  \\
   T2M3 ..... &  $10^{-2}$ & $3\times 10^{-2}$   &  $2\times 10^{-3}$       &  $10^{-6}$     &  27    \\
   T2M1 ..... &  $10^{-2}$ & $3\times 10^{-2}$   &  $2\times 10^{-1}$       &  $10^{-5}$     &  2.7\\
   \hline
  \end{tabular}
 Note.$-$ The dimensionless parameter $\mu_{\rm 0}$ indicates the mass-to-flux ratio normalized by the critical value $(M/\Phi)_{\rm cr}$.
\end{table}
\begin{figure*}
\includegraphics[trim=30 10 20 0, scale = 0.78, clip]{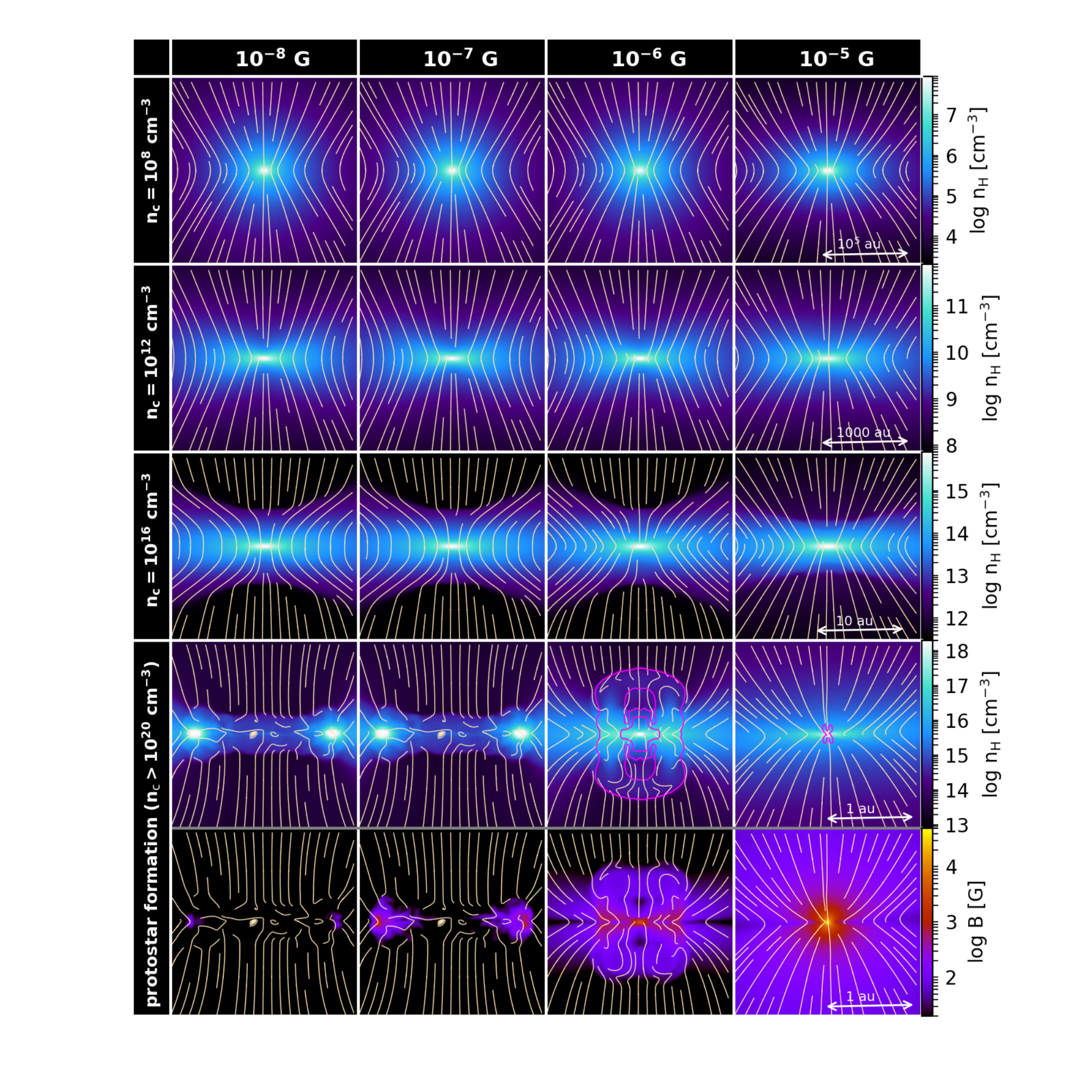} 
\caption{
The edge-on sliced density distributions when the central density $n_{\rm c}= 10^{8}\ \rm cm^{-3},$ $10^{12}\ \rm cm^{-3},$ $10^{16}\ \rm cm^{-3}$ and just after protostar formation 
for pure rotation cases, with four different strengths of initial magnetic field $B_{\rm init}=10^{-8},\ 10^{-7},\ 10^{-6},\ 10^{-5}\ \rm G$ (from the left to right columns).
The bottom row indicates the magnetic field distribution at the epoch of protostar formation.
Magnetic field lines projected on the plane are drawn by white lines in each panel.
Outflow regions, characterised by outward radial velocities, are encircled by the magenta lines in the right two columns.
}
\label{no_turb_snap}
\end{figure*}

In the turbulent cases, how the small-scale dynamo proceeds depends on the ratio of magnetic and turbulent energies, as seen in Section \ref{sec_dynamo}.
Therefore, we classify the simulation runs of turbulent cases into three cases according to their initial ratios as follows:
\begin{itemize}
\item {\it super-Alfv\'{e}nic case}; 
the runs with $B_{\rm init}=10^{-9}, 10^{-8}$, and $10^{-7}\ \rm{G}$, where the turbulent energy is sufficiently larger than the magnetic energy ($E_{\rm mag}/E_{\rm turb}\simeq 7\times10^{-8}, 7\times10^{-6}$, and  $7\times10^{-4}$, respectively)

\item {\it trans-Alfv\'{e}nic case}; 
the run with $B_{\rm init}=10^{-6}\ \rm{G}$ $(E_{\rm mag}/E_{\rm turb}\simeq7\times10^{-2})$, where
the magnetic energy is roughly equal to the turbulent energy

\item {\it sub-Alfv\'{e}nic case};
the run with $B_{\rm init}= 10^{-5}\ \rm{G}$ 
$(E_{\rm mag}/E_{\rm turb}\simeq 7\times10^{0})$,
where the magnetic energy is sufficiently larger than the turbulent energy.
\end{itemize}
\begin{figure*}
\includegraphics[trim=30 150 10 150, scale = 0.68, clip]{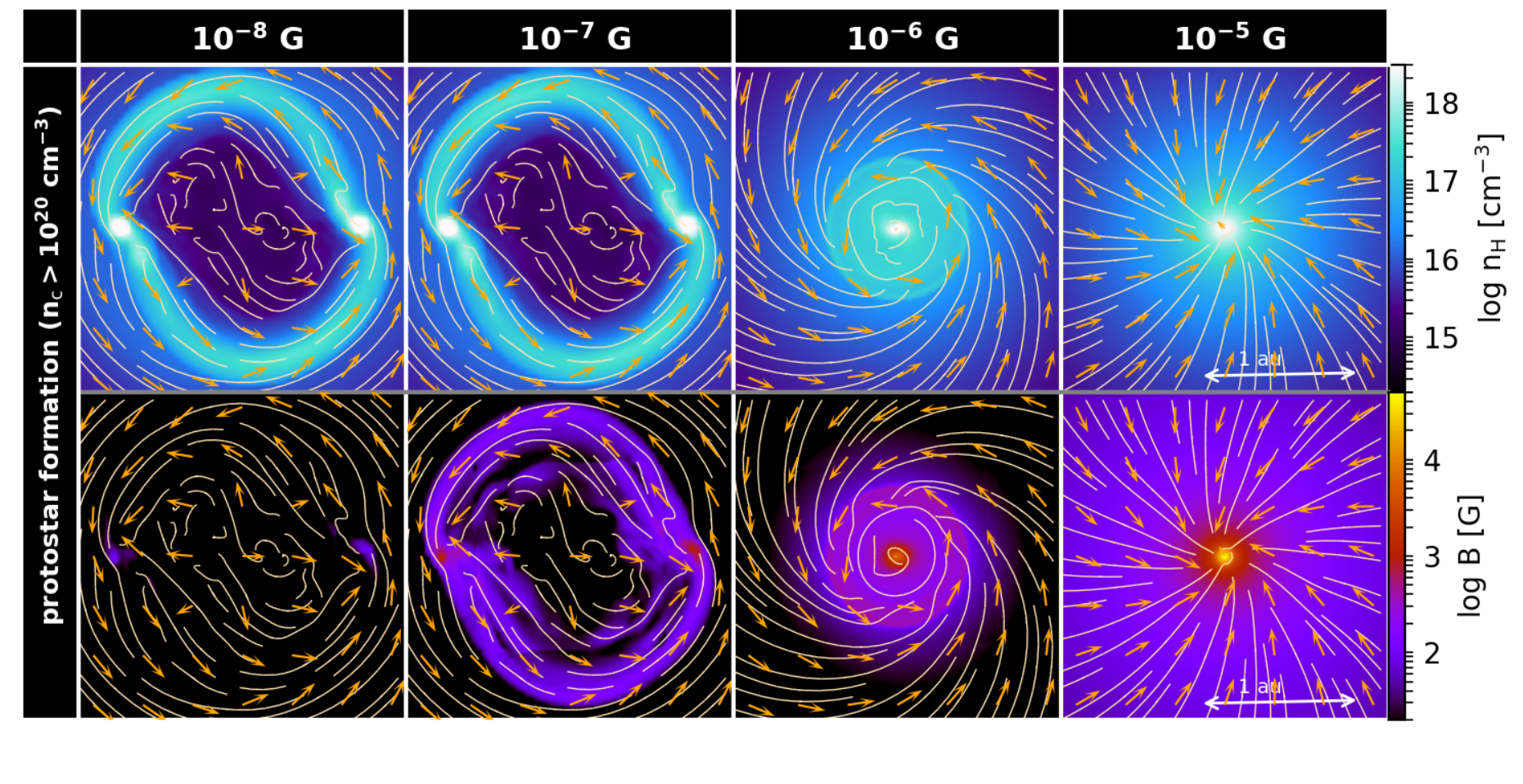} 
\caption{
The density (top) and magnetic field (bottom) distributions (face-on view) at the epoch of protostar formation for a cloud initially only with rotational motion and with four different strengths of the initial magnetic field $B_{\rm init}=10^{-8},\ 10^{-7},\ 10^{-6},\ 10^{-5}\ \rm G$ (from the left to right columns).
Magnetic field lines projected on the plane are drawn by white lines in each panel.
The direction of the velocity is indicated by the orange arrows. 
}
\label{snap_face_on}
\end{figure*}
\begin{figure}
\includegraphics[trim=0 70 20 80, scale = 0.44, clip]{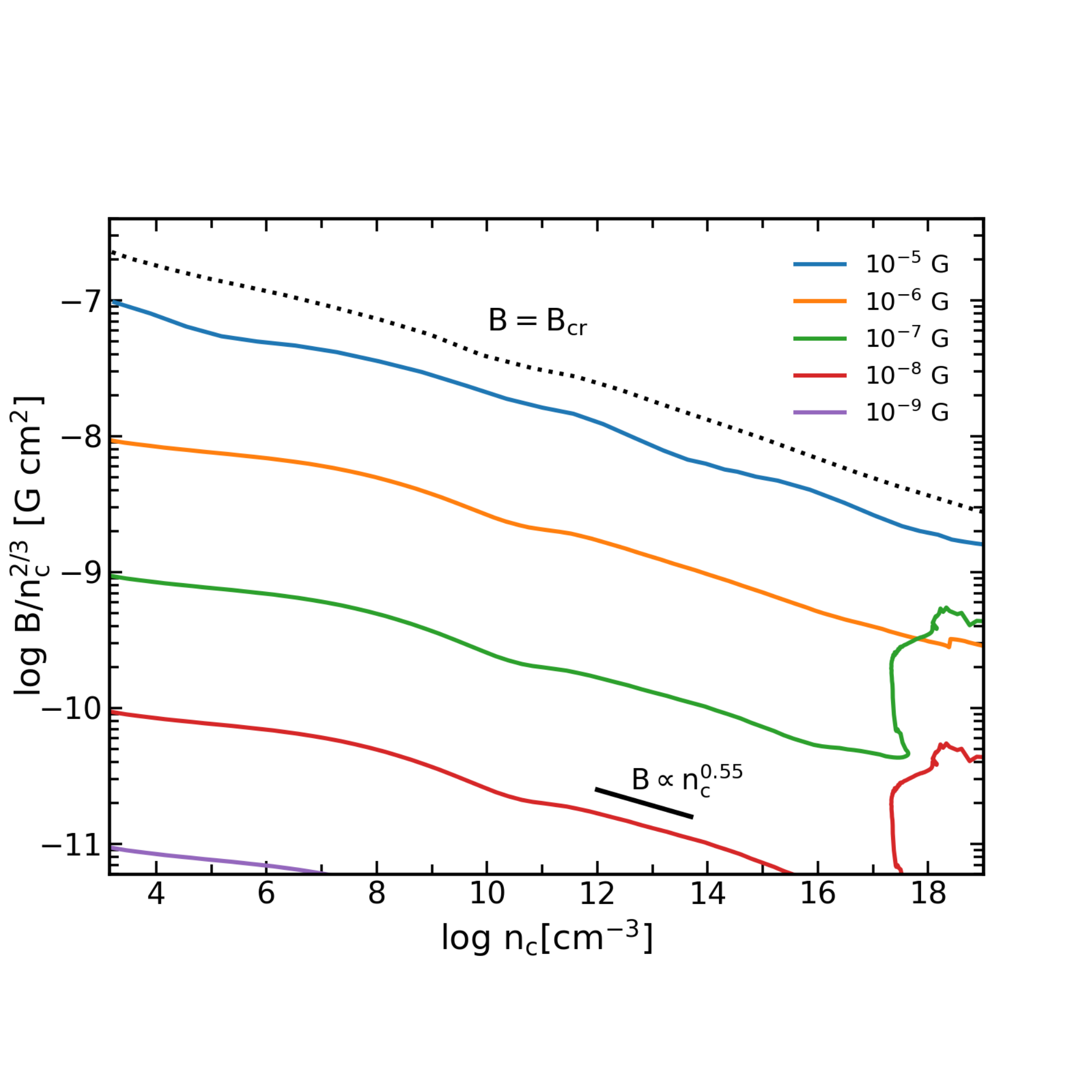}
\caption{
The evolution of magnetic field at the centre for cases only with rotation at the beginning.
Shown is $B/n^{2/3}_{\rm c}$, which becomes constant for the amplification by spherical collapse, as a function of the central density $n_{\rm c}$.
The colors show the different initial magnetic field values $B_{\rm init}= 10^{-8}\ \rm G$ (red), $10^{-7}\ \rm G$ (green), $10^{-6}\ \rm G$ (orange), and $10^{-5}\ \rm G$ (blue), respectively.
The critical field strength (eq. \ref{eqBcr}) is also indicated by the black dotted line.
The thick black line represents the field amplification rate in the sheet-like collapse for $T_{\rm c}\propto n_{\rm c}^{\gamma_{\rm eff}-1}$($\gamma_{\rm eff}\simeq1.1$)
}
\label{rot_mag_evo}
\end{figure}
\section{Result}\label{sec_result}
In this section, we first examine the pure rotation cases with AD in Sec. \ref{sec_rot} to see how the magnetic field is amplified during the collapse and how strong field is required for causing such dynamical effects as magnetic braking or MHD outflow launching.
Next, we move on to the turbulent cases with AD in Sec. \ref{sec_turb}, and compare them with the pure rotation cases. 
We then discuss the influence of AD in Sec. \ref{sec_AD} from the comparison between the runs with and without AD.

\begin{figure*}
\includegraphics[trim=30 30 20 0, scale = 0.78, clip]{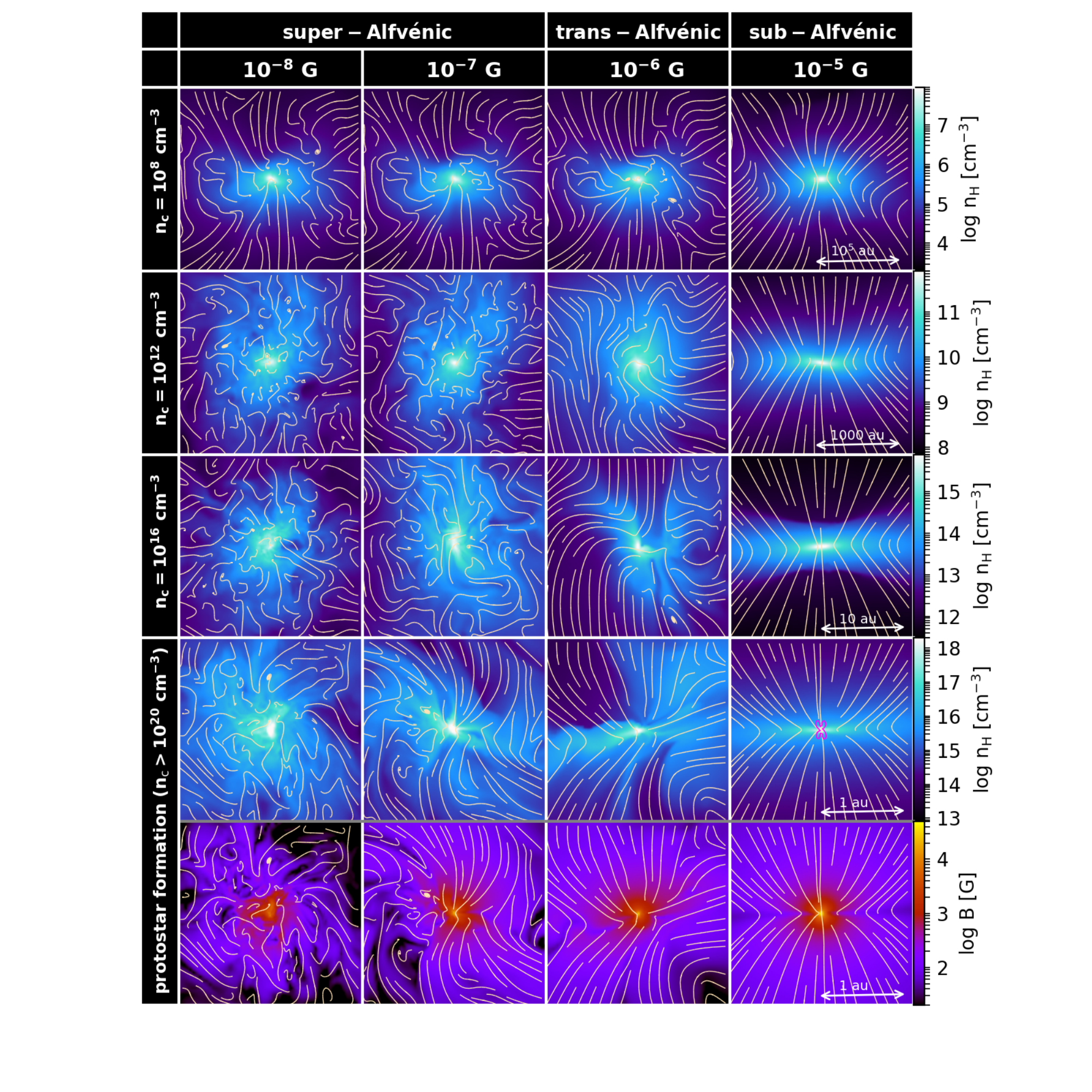} 
\caption{ 
The same as Fig. \ref{no_turb_snap}, but for the turbulent cases.
}
\label{turb_snap}
\end{figure*}
\subsection{Pure rotation cases}\label{sec_rot}
We show in Fig. \ref{no_turb_snap} the collapse of a cloud initially with pure rotation up to the protostar formation for the cases with different initial field strengths $B_{\rm init} = 10^{-8},\ 10^{-7},\ 10^{-6},$ and $10^{-5}\ \rm G$.
The four rows from the top show the edge-on sliced density distributions at four different epochs with the central densities $n_{\rm c} = 10^{8},\ 10^{12},\ 10^{16}\ \rm cm^{-3},$ and of the protostar formation ($n_{\rm c}>10^{20}\ \rm cm^{-3}$).
For the last epoch, the magnetic field strength is also shown in the bottom panels.
Projected field lines on each plane are represented with white lines.
Although we have performed the simulations with the AD effect in a similar setting to previous work (\citealp{Sadanari2021}),
its effect is not significant in the pure rotation cases, as we will see below (also see the discussion in Section \ref{sec_AD}).

In all the cases, the clouds collapse in a runaway fashion where the density in the central region increases on a free-fall timescale $t_{\rm ff}=\sqrt{3\pi/32 G \rho }\propto \rho^{-1/2}$, with a lower-density outer region being left behind.
The resulting density profile consists of a constant density central core of roughly the Jeans length $L_{\rm J}=c_{s}\sqrt{\pi/G\rho}$, where $c_{s}$ represents the sound velocity at the center,
and an envelope with a power-low density distribution $\rho \propto r^{-2.2}$ (\citealp{Omukai_Nishi1998}).

As seen in Fig. \ref{no_turb_snap}, the shape of the cloud changes from spherical to elliptical as the collapse proceeds, and finally transforms into disc-like in all the cases.
In the weak field cases of $B_{\rm init} \leq 10^{-6}\rm \ G$, this is mainly due to the centrifugal forces, 
while in the case with the strongest field of $B_{\rm init}=10^{-5}\ \rm G$, it is the magnetic force that deforms the cloud. 
Such disc-like structure supported by anisotropic magnetic tension, rather than rotation, is a pseudo-disc (\citealp{Galli_Shu1993}).

Whether the cloud eventually fragments to form a multi-protostar system is determined by the initial field strength relative to the rotation.
Fig. \ref{snap_face_on} shows the density and magnetic field distributions on the face-on view at the protostar formation epoch for four different $B_{\rm init}$ cases. The orange arrows and white lines indicate the projected velocity direction and field lines, respectively.
In the cases with a weak initial magnetic field of $B_{\rm init} \leq 10^{-7}\ \rm G$ ($E_{\rm mag}/E_{\rm rot}\leq 2\times10^{-3})$,
the rotating disc gradually transforms into a ring-like structure of $\sim 1\ \rm au$ by the centrifugal force.
Since the ring is gravitationally unstable, it finally breaks up into a binary system.
With a stronger field of $B_{\rm init}=10^{-6}\ \rm G$ ($E_{\rm mag}/E_{\rm rot}= 2\times10^{-1}$), 
the size of the rotating disc decreases to $\simeq 0.5 \ \rm au$ 
due to the angular momentum transport by magnetic braking and outflow launching.
Such a small disc is gravitationally stable against fragmentation and just one protostar is formed at the center.
In the strongest field case of $B_{\rm init}=10^{-5}\ \rm G$ ($E_{\rm mag}/E_{\rm rot}= 2\times10^{1}$), most of the angular momentum is extracted immediately after the onset of the collapse. 
Thus, one protostar is formed at the center of the hardly rotating pseudo-disc (see orange arrows in the rightmost column of Fig. \ref{snap_face_on}).
To summarize, the fragmentation occurs when the initial magnetic energy $E_{\rm mag}$ is considerably smaller than the rotational energy $E_{\rm rot}$.
This fragmentation condition is the same as that found in ideal-MHD simulations (\citealp{Machida2008}; \citealp{Sadanari2021}), 
suggesting that AD does not affect the efficiency of magnetic braking in the pure rotation cases.

The outflow regions, where the radial velocity $V_{\rm r}$ is outward, are indicated by magenta lines in the fourth row from top of Fig. \ref{no_turb_snap}.
The outflows are launched in the strong field cases of $B_{\rm init}=10^{-6}\ \rm G$ and $10^{-5}\ \rm G$,
with different launching mechanisms.
In the case of $B_{\rm init}=10^{-5}\ \rm G$, the magneto-centrifugal wind is
 driven by the centrifugal force along field lines (\citealp{Blandford_Payne1982}; \citealp{Tomisaka2002}; \citealp{Machida2008a}), as suggested from the predominant poloidal field configuration (see the field line in Fig. \ref{no_turb_snap} and Fig. \ref{snap_face_on}). 
By contrast, in the case of $B_{\rm init}=10^{-6}\ \rm G$, the magnetic-pressure wind (\citealp{Tomisaka2002}; \citealp{Banerjee_Pudritz2006}; \citealp{Machida2008a}; \citealp{Tomida2013}) is launched owing to  outwardly decreasing magnetic pressure gradient (bottom row of Fig. \ref{no_turb_snap}), 
 created by the amplification of a toroidal field by the rotating disc (see the third panel of Fig. \ref{snap_face_on}).

Next, we focus on the field amplification during the collapse.
In Fig. \ref{rot_mag_evo}, 
we plot the mass-weighted average of the normalized field strength over the central region.
Here, the central region is defined as the region within the Jeans radius $R_{\rm J}=L_{\rm J}/2$.
The abscissa and ordinate indicate the average number density in the central region $n_{\rm c}$ and the normalized field strength $B/n^{2/3}_{\rm c}$, respectively.
We also plot the critical field $B_{\rm cr}$ (black dotted) as an indicator of whether the magnetic force affects the gas dynamics, 
i.e., the closer the field strength approaches $B_{\rm cr}$, the more the field affects the dynamics.
The critical field $B_{\rm cr}$ is defined as the field strength at which the magnetic force ($\propto |(\nabla \times B)\times B|$) and gravity of a uniform density core with the Jeans radius $R_{\rm J}$ balance. 
This can be written as
\begin{equation} \label{eqBcr}
B_{\rm cr} = \sqrt{\frac{4\pi G M_{\rm J}\rho_{\rm c}}{R_{\rm J}}},
\end{equation}
where $\rho_{\rm c}$ and $M_{\rm J}\propto \rho_{\rm c}R^3_{\rm J}$ are the average mass density and the mass of the central core, respectively.
Note that the central magnetic field during the collapse cannot exceed $B_{\rm cr}$.

In the absence of turbulence, the central field is mostly amplified by the global compression associated with gravitational collapse.
In this case, the field amplification rate depends on the cloud morphology.
The magnetic field increases as $B \propto n_{\rm c}^{2/3}$ in the spherical collapse
and as $B \propto n_{\rm c}^{1/2}$ in the sheet-like collapse
\footnote{
The relationship between the magnetic field strength and the gas density can be derived from the mass and flux conservation law.
Here, the central mass density $\rho_{\rm c}$ and the magnetic field strength $B$ are assumed to be constant in the central core of radius $R_{\rm c}$.
In the case of spherical collapse, the mass and flux conservation are $\rho_{\rm c}R_{\rm c}^3=\rm const.$ and  $BR_{\rm c}^2=\rm const.$, respectively.
From these relations, we can derive $B\propto n_{\rm c}^{2/3}$.
Similarly, in the case of sheet-like collapse with the scale height $H=c_{\rm s}/\sqrt{G\rho_{\rm c}}\propto n^{-1/2}_{\rm c}$, 
the mass and flux conservation are $\rho_{\rm c}R_{\rm c}^2H=\rm const.$ and $BR_{\rm c}^2=\rm const.$, respectively,  and thus $B \propto n_{\rm c}^{1/2}$.
}.
As seen in Fig. \ref{no_turb_snap}, the cloud immediately becomes somewhat disc- or sheet-like in all the cases either by the centrifugal or magnetic forces, 
so that the central field grows more like $B \propto n^{1/2}_{\rm c}$ rather than $\propto n^{2/3}_{\rm c}$ (Fig. \ref{rot_mag_evo}).
\footnote{
More precisely, the field grows as $B\propto n_{\rm c}^{\gamma_{\rm eff}/2}\simeq n_{\rm c}^{0.55}$ (see the thick black line in Fig. \ref{rot_mag_evo}), reflecting the increasing temperature with the effective ratio of specific heat $\gamma_{\rm eff} \simeq 1.1$, i.e., $T_{\rm c} \propto n_{\rm c }^{0.1}$, for the primordial gas.
}
Since the critical field $B_{\rm cr}$ has the same density dependence ($B_{\rm cr} \propto n^{1/2}_{\rm c}$ for constant temperature), the gravitational compression amplification alone cannot amplify the weak field to a level close to $B_{\rm cr}$.
As a special case, when the ring is formed and the collapse is temporarily suppressed ($B_{\rm init}\leq 10^{-7}\ \rm G$), 
the field can also be amplified by the stretching of field lines due to rotational motion.
This can be seen as the almost vertical jump of the field strength around $n_{\rm c}\sim 10^{18}\ \rm cm^{-3}$ in the cases of $B_{\rm init} = 10^{-8}\ \rm G$ (red) and $10^{-7}\ \rm G$ (green) in Fig. \ref{rot_mag_evo}.
Since this occurs just before the protostar formation, 
no significant dynamical effect is observed in those cases.
To summarize, magnetic fields must be close to $B_{\rm cr}$ from the beginning to have a significant dynamical effect on the collapsing clouds in the pure rotation cases.
The situation will change, however, in the presence of turbulence, as we will see below. 

\begin{figure}
\includegraphics[trim=0 70 20 80, scale = 0.44, clip]{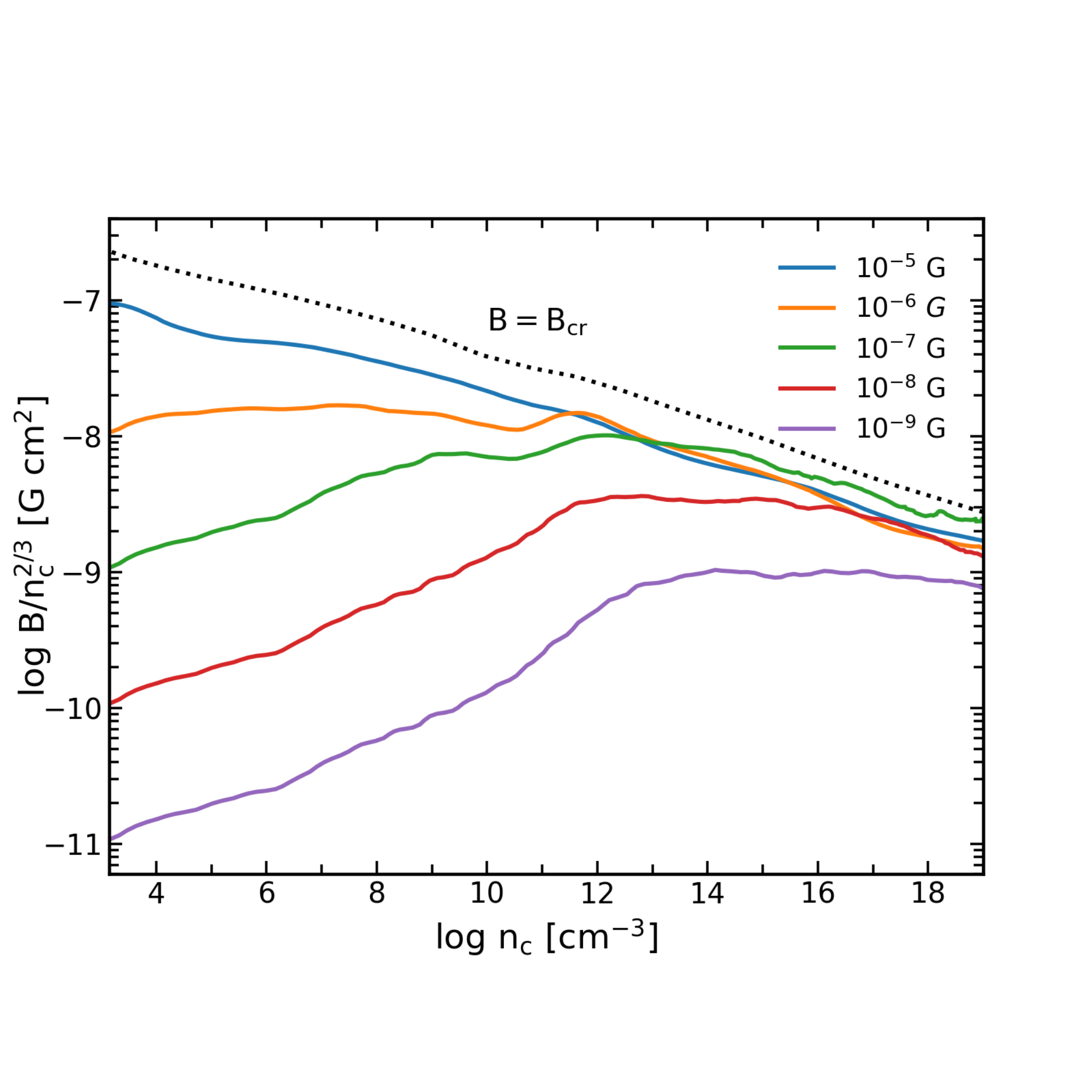}
\caption{
The same as Fig. \ref{rot_mag_evo}, but for the turbulent cases.
In the super-Alfv\'{e}nic cases with $B_{\rm init}=10^{-9}\ \rm G$ (purple), $10^{-8}\ \rm G$ (red), and $10^{-7}\ \rm G$ (green), 
the magnetic fields are amplified rapidly by the kinematic dynamo in the early stage; 
in the trans-Alfv\'{e}nic case with $B_{\rm init}=10^{-6}\ \rm G$ (orange) 
the magnetic field is amplified slowly by the non-linear dynamo from the beginning;
and in the sub-Alfv\'{e}nic case with $B_{\rm init}=10^{-5}\ \rm G$ (blue), the dynamo amplification is not observed.
}
\label{mag_evo}
\end{figure}
\begin{figure*}
\hspace*{-0.8cm}
\includegraphics[trim=0 200 0 220, scale = 0.95, clip]{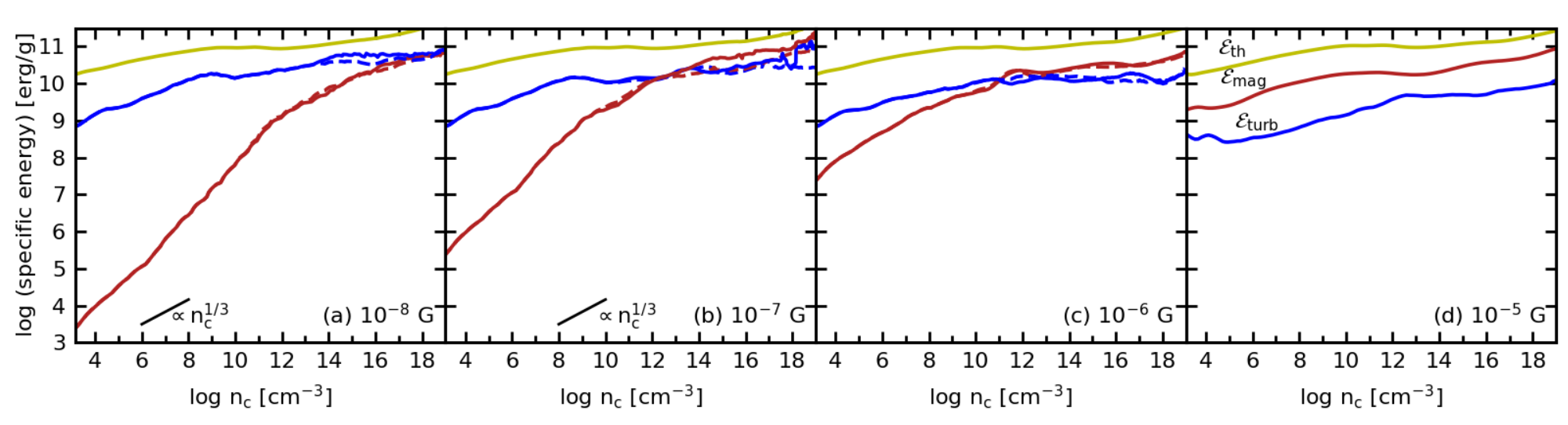}
\caption{
The core-averaged energy densities of different components: magnetic energy $\mathcal{E}_{\rm mag}=B^2/(8\pi \rho)$ (red), turbulent energy $\mathcal{E}_{\rm turb}=V^2_{\rm turb}/2$ (blue), and thermal energy $\mathcal{E}_{\rm th}=3c^2_{\rm s}/2$ (yellow), as a function of the central density $n_{\rm c}$ for the initial field strengths (a) $B_{\rm init} = 10^{-8}$, (b) $10^{-7}$, (c) $10^{-6}$, and (d) $10^{-5} \ {\rm G}$.
The cases with and without AD are shown by solid and dashed lines, respectively.
The thick black lines in Figs. \ref{cento_Eevo}(a) and \ref{cento_Eevo}(b) represent the field amplification rate in the spherical compression, 
i.e., $\mathcal{E}_{\rm mag}\propto B^{2}/n_{\rm c} \propto n_{\rm c}^{1/3}$.
The two lines for the thermal energy (in all the cases) and the other energies (in the cases of $B_{\rm init} = 10^{-5} {\rm G}$)
completely overlap with each other. 
}
\label{cento_Eevo}
\end{figure*}
\begin{figure*}
\hspace*{-0.8 cm}
\includegraphics[trim=0 142 0 280, scale = 0.95, clip]{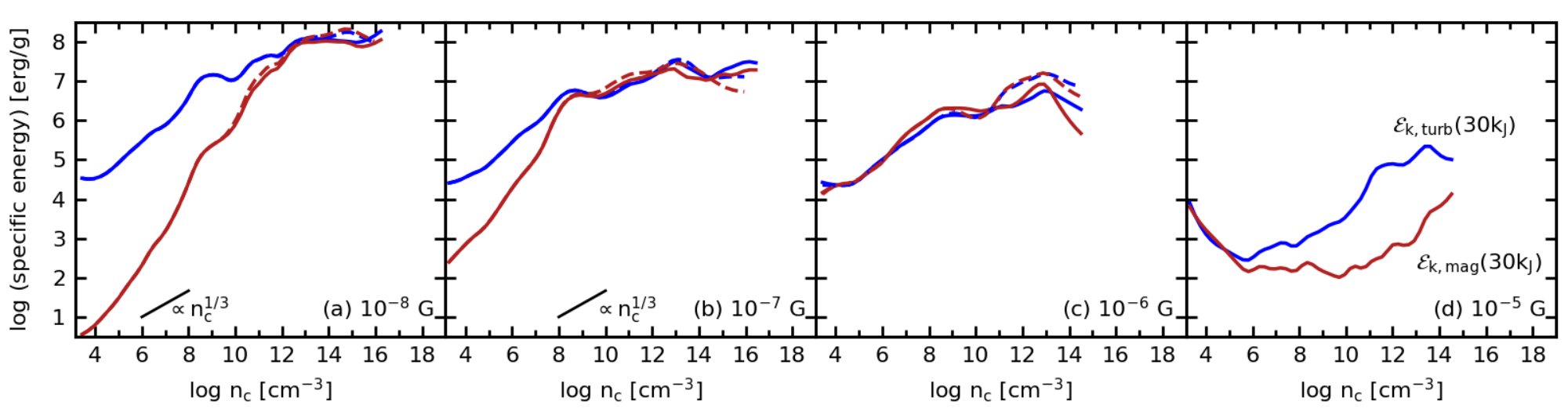} 
\caption{
The same as Fig. \ref{cento_Eevo}, but for the magnetic and turbulent energy densities on the scale $k=30k_{\rm J}$ (the smallest scale in the current simulations).
The point at which the magnetic energy (red) catches up with the turbulent energy (blue) corresponds to the transition point from the kinematic to the non-linear stages in the case of $B_{\rm init}=$ (a) $10^{-8}\ \rm G$ and (b) $10^{-7}\ \rm G$. In the case of (c) $10^{-6}\ \rm G$, the two energy values are equal, and thus the dynamo is already in the non-linear stage from the beginning.
}
\label{cento_Eevo30}
\end{figure*}
\begin{figure}
\includegraphics[trim=130 0 30 20, scale = 0.73, clip]{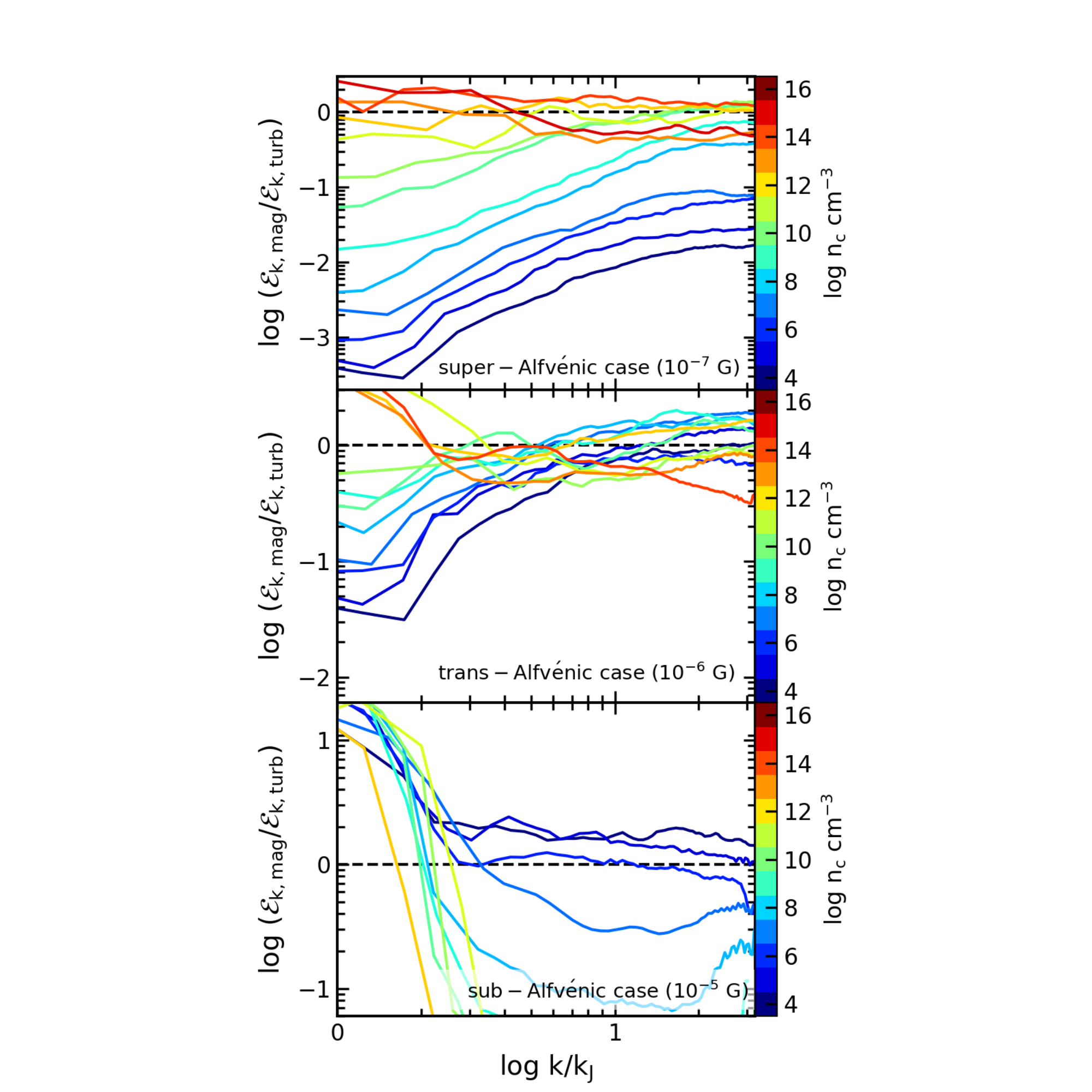} 
\caption{
Evolution of the ratio of the magnetic to turbulent energy spectra
for the super-Alfv\'{e}nic case of $B_{\rm init}=10^{-7}\ \rm G$ (top panel), 
the trans-Alfv\'{e}nic case of $B_{\rm init}=10^{-6}\ \rm G$ (middle panel), 
and the sub-Alfv\'{e}nic case of $B_{\rm init}=10^{-5}\ \rm G$ (bottom panel). 
The colors indicate the central density when the spectra are taken
($n_{\rm c}=10^{4},\ 10^{5},\cdots ,\ 10^{16}\ \rm cm^{-3}$ from dark blue to dark red).
Horizontal axis is the wavenumber normalized by the local Jeans wavenumber $k_{\rm J}(=1/L_{\rm J})$ at each time.
}
\label{sepect_64}
\end{figure}
\subsection{Turbulent cases}\label{sec_turb}
Next, we discuss the cases with turbulence. 
We show the evolution with four different initial fields $B_{\rm init}=10^{-8},\ 10^{-7},\ 10^{-6},$ and $\ 10^{-5}\ \rm G$ in Fig. \ref{turb_snap}, as in Fig. \ref{no_turb_snap}. 
Here, we classify those cases into three groups as mentioned in Sec. \ref{sec_init}:
super-Alfv\'{e}nic cases ($B_{\rm init}\leq 10^{-7}\ \rm G$), trans-Alfv\'{e}nic case ($B_{\rm init}=10^{-6}\ \rm G$), and sub-Alfv\'{e}nic case ($B_{\rm init}=10^{-5}$).
In Fig. \ref{turb_snap}, 
we can see that that turbulence is suppressed by the presence of a strong coherent field in the sub-Alfv\'{e}nic case, and thus the cloud evolves in the same manner as in the pure rotation case (Fig. \ref{no_turb_snap}). 
By contrast, turbulence survives in the trans- and super-Alfv\'{e}nic cases,
and the cloud morphology and magnetic field lines are disturbed by the turbulent motion.
Especially in the super-Alfv\'{e}nic cases ($B_{\rm init} \leq 10^{-7}\ \rm G$),
unlike in the pure rotation cases, the cloud collapses spherically on average, forming a single protostar at the centre.
Since the direction of mean angular momentum in the central core varies significantly during the collapse, 
we infer that shear motion of turbulence can induced the angular momentum transport (e.g., \citealp{Greif2012}).

As noted in Sec. \ref{sec_dynamo}, 
turbulence can drive the small-scale dynamo by stretching and folding the field lines, 
and amplify magnetic fields more efficiently than with the gravitational compression alone.
We can see that the dynamo is operating in the super- and trans-Alfv\'{e}nic cases from the fact that the field lines are disturbed randomly (Fig. \ref{turb_snap}).
As a result, unlike the cases with pure rotation, 
even an initially weak field can be amplified to the same level as in the strongest field case ($B_{\rm init}=10^{-5}\ \rm G$) by the epoch of protostar formation
 (compare the bottom panels of the Fig. \ref{no_turb_snap} and Fig. \ref{turb_snap}).
Note that, in our calculations, the magnetic field strength around the protostar reaches $B=10^{3}$-$10^{5}\ \rm G$, more than two orders of magnitude larger than
found in present-day star formation simulations (e.g., \citealp{Machida2007}; \citealp{Vaytet2018}; \citealp{Machida_Basu2019}; \citealp{Wurster2022}). 
This difference comes from the fact that the field dissipates more efficiently in the present-day case, via the Ohmic dissipation and AD, due to the lower ionisation degree thanks to the presence of dust.
This suggests that AD in the primordial gas is not efficient enough to inhibit the field amplification (see Sec. \ref{sec_AD} for details).

To see the effect of dynamo amplification, we plot the central value of the normalized magnetic field $B/n^{2/3}_{\rm c}$ in Fig. \ref{mag_evo} in the same manner as in Fig.\ref{rot_mag_evo}.
Recall that the global compression can amplify the field as $B\propto n^{2/3}_{\rm c}$ at most.
Hence, any increase of normalized field $B/n^{2/3}_{\rm c}$ with density in Fig. \ref{mag_evo} can be ascribed to the kinematic dynamo amplification (Sec. \ref{sec_dynamo}).
This is indeed observed in the super-Alfv\'{e}nic cases ($B_{\rm init}\leq 10^{-7}\ \rm G$).
For the trans-Alfv\'{e}nic case ($B_{\rm init}=10^{-6}\ \rm G$, orange), the normalized field strength remains constant, suggesting that the amplification by the global compression associated with spherical collapse (in the averaged sense) is dominant over the dynamo effect, although the field orientation is highly disturbed by the turbulence (Fig. \ref{turb_snap}).
This indicates that the dynamo amplification is already in the non-linear stage, in which the field growth is inefficient due to the back-reaction of magnetic forces (Sec. \ref{sec_dynamo}).
In the sub-Alfv\'{e}nic case ($B_{\rm init}=10^{-5}\ \rm G$, blue), the turbulence is swiftly suppressed by the strong coherent magnetic field as seen in Fig. \ref{turb_snap},
and thus the field is amplified only by the global compression.

Below, we examine how the dynamo amplifies the magnetic field in more detail focusing both on core-averaged and scale-dependent quantities.
To this end, here we introduce the 1D power spectrum $E_{k,\rm mag/turb}$ defined as the spherical shell average of the 3D power spectrum in the wavenumber $k$-space, as
\begin{equation} 
\label{eq_Ekmag}
\mathcal{E}_{\rm mag} = \frac{1}{k_{\rm J}^3}\int^{\infty}_{k_{\rm J}} \frac{\langle |\hat{B}(\bm{k})|^2\rangle}{8\pi \rho_{\rm c}}4\pi k^2 dk = \int^{\infty}_{k_{\rm J}} E_{k, \rm mag}(k)dk,
\end{equation}
\begin{equation} 
\label{eq_Ekturb}
\mathcal{E}_{\rm turb} = \frac{1}{k_{\rm J}^3}\int^{\infty}_{k_{\rm J}}\frac{\langle |\hat{V}_{\rm turb}(\bm{k})|^2\rangle}{2} 4\pi k^2 dk= \int^{\infty}_{k_{\rm J}} E_{k, \rm turb}(k)dk,
\end{equation}
where $\hat{B}(\bm{k})$ and $\hat{V}_{\rm turb}(\bm{k})$ are the Fourier components of magnetic field and the turbulent velocity, respectively,
and $k_{\rm J}$ is the local Jeans wavenumber, i.e., $k_{\rm J}=1/L_{\rm J}$.
We also define $\mathcal{E}_{k,\rm mag}(k) \equiv kE_{k, \rm mag}$ and $\mathcal{E}_{k,\rm turb}(k) \equiv kE_{k, \rm turb}$ as the specific energy densities on the scale $k$.  
With the scale-dependent specific energy densities defined above, we will see three different quantities: 
the core-averaged energy density, 
 smallest-scale energy density, and scale-dependence of energy density ratio.
Firstly, Fig. \ref{cento_Eevo} shows 
the core-averaged magnetic field energy density $\mathcal{E}_{\rm mag}=\langle B^2/(8\pi \rho) \rangle$ (red), turbulence energy density $\mathcal{E}_{\rm turb}=\langle V^2_{\rm turb}/2\rangle$ (blue), 
and thermal energy density $\mathcal{E}_{\rm th}=\langle3c^2_{\rm s}/2\rangle $ (yellow) for the cases of $B_{\rm init}=$ (a) $10^{-8}\ \rm G$, (b) $10^{-7}\ \rm G$, (c) $10^{-6}\ \rm G$, and (d) $10^{-5}\ \rm G$.
Here, we define the turbulent velocity $V_{\rm turb}$ as the remainder after the subtraction of the shell-averaged radial velocity component.
Note that the rotational component has not been excluded for the simplicity of the analysis.
The solid and dashed lines represent the cases with and without AD, respectively, and the differences between them will be discussed in Sec. \ref{sec_AD}.
Next, in Fig. \ref{cento_Eevo30}, we plot the evolution of the magnetic and turbulent energy density on the smallest scale in our simulations, i.e., $k=30k_{\rm J}$, as in Fig. \ref{cento_Eevo}.
We chose $30k_{\rm J}$ from the observation that the power spectra of $E_{k,\rm mag}$ and $E_{k,\rm turb}$ decay rapidly below this scale due to the numerical dissipation.
Finally, we show the evolution of the ratio of the magnetic to turbulent energy spectra $\mathcal{E}_{k,\rm mag}/\mathcal{E}_{k, \rm turb}$ as a function of the normalized wavenumber $k/k_{\rm J}$ in Fig. \ref{sepect_64}.
The top, middle, and bottom panels show the cases of $B_{\rm init}=10^{-7}\ \rm G$, $10^{-6}\ \rm G$, and $10^{-5}\ \rm G$, respectively.
The colors correspond to the epochs labelled by the central density $n_{\rm c}$.
From now on, with Figs. \ref{cento_Eevo}, \ref{cento_Eevo30}, and \ref{sepect_64}, we carefully examine the magnetic field amplification in the super-, trans- and sub-Alfv\'{e}nic cases in this order.

Firstly, in the super-Alfv\'{e}nic cases ($B_{\rm init}=10^{-8}\ {\rm and}\ 10^{-7}\rm G$), having smaller energy compared with the turbulence,
the magnetic fields on each scale, especially on the smaller scales, rapidly grow with the eddy turnover timescale $t_{\rm eddy}(k)$ by the kinematic dynamo (Figs. \ref{cento_Eevo30}a and \ref{cento_Eevo30}b).
As a result, the amplification of the core-averaged magnetic energy (red line in Figs. \ref{cento_Eevo}a and \ref{cento_Eevo}b) exceeds that by the spherical compression, i.e.,  $\mathcal{E}_{\rm mag} \propto B^2/n_{\rm c} \propto n_{\rm c}^{1/3}$. Note that the collapse proceeds in a roughly spherical fashion (see Fig. \ref{turb_snap}, left two columns).
As expected for the kinematic dynamo (Sec. \ref{sec_dynamo}),
the magnetic energy on the smaller scale $\mathcal{E}_{k,\rm mag}(k)$ approaches the turbulent energy $\mathcal{E}_{k,\rm turb}(k)$ faster in the evolution (Fig.\ref{sepect_64}, top).

The kinematic stage comes to an end and the non-linear stage begins when the magnetic field energy becomes comparable to the turbulence energy on the smallest scale in the simulations, 
i.e., $\mathcal{E}_{k,\rm mag}(30k_{\rm J})\sim \mathcal{E}_{k,\rm turb}(30k_{\rm J})$.
From Fig. \ref{cento_Eevo30}, we can clearly identify the transition occurring at $n_{\rm c}\sim 10^{12}\ \rm cm^{-3}$ for the case of $B_{\rm init}=10^{-8}\ \rm G$ (Fig. \ref{cento_Eevo30}a) and at $n_{\rm c}\sim 10^{9}\ \rm cm^{-3}$ for the case of $B_{\rm init}=10^{-7}\ \rm G$ (Fig. \ref{cento_Eevo30}b), respectively.
After entering the non-linear stage, 
the dynamo amplification on the small scales where the equipartition is achieved, is hindered by the back-reaction of magnetic forces,
while on the larger scales where the equipartition has not been reached yet, the magnetic field continues to grow rapidly without back-reaction effect (Fig. \ref{sepect_64}, top).
Accordingly, the range of equipartition extends toward a larger scale (Fig. \ref{sepect_64}, top).
During the non-linear stage, the core-averaged field energy grows only linearly with time ($\mathcal{E}_{\rm mag}\propto t$) by the non-linear dynamo,
and thus the amplification is dominated by the global compression, i.e., $\mathcal{E}_{\rm mag} \propto n^{1/3}_{\rm c}$.

The non-linear stage of the turbulent dynamo ends when the field reaches equipartition on the largest turbulent-driving scale ($\sim k_{\rm J}$), 
corresponding to when $\mathcal{E}_{\rm mag}\sim \mathcal{E}_{\rm turb}$ in the core-averaged sense is achieved.
In the case of $B_{\rm init}=10^{-7}\ \rm G$, we see that this occurs at $n_{\rm c} \sim 10^{12}\ \rm cm^{-3}$ (Fig. \ref{cento_Eevo}b).
After that, the magnetic field is amplified only by the global gravitational compression, resulting in a generation of the coherent field on the Jeans scale,
as we can see in Fig. \ref{turb_snap} that the field lines roughly align each other at the protostar formation in the case of $B_{\rm init}=10^{-7}\ \rm G$.
Once a strong coherent field is established, it begins to suppress turbulent motion, so that the turbulent energy in Fig. \ref{cento_Eevo}(b) becomes lower than the magnetic field energy ($n_{\rm c}>10^{12}\ \rm cm^{-3}$). Consistently, the field energy on the Jeans scale exceeds the turbulent energy, i.e., $\mathcal{E}_{k,\rm mag}(k_{\rm J})> \mathcal{E}_{k,\rm turb}(k_{\rm J})$ in Fig. \ref{sepect_64} (top).  
Since $\mathcal{E}_{\rm mag}, \mathcal{E}_{\rm turb},$ and $\mathcal{E}_{\rm th}$ are all in the same order of magnitude during this epoch (Fig. \ref{cento_Eevo}b),
$B$, $B_{\rm eq}$ and $B_{\rm cr}$ are also in the same order of magnitude (see eqs. \ref{eq_Beq} and \ref{eqBcr}), indicating that the saturated magnetic field has a potential to affect the gas dynamics.
In the case of $B_{\rm init}=10^{-8}\ \rm G$, the non-linear stage ends at $n_{\rm c}\sim 10^{18}\ \rm cm^{-3}$ (Fig. \ref{cento_Eevo}a),
which is too late for the gravitational compression to generate a coherent field before the protostar formation.
As a result, the field configuration at the protostar formation is more random compared to the case of $B_{\rm init}=10^{-7}\ \rm G$ (Fig. \ref{turb_snap}, bottom and second from the bottom).

Next, let us examine the trans-Alfv\'{e}nic case ($B_{\rm init}=10^{-6}\ \rm G$). 
As seen from Fig. \ref{cento_Eevo30}c,  the magnetic energy is comparable to the turbulent energy on the smallest scale $k\sim 30k_{\rm J}$ from the beginning, 
suggesting that the turbulence is affected by the field back-reaction especially on the small scales.
Therefore, the range of equipartition with the turbulence extends to a larger scale (Fig. \ref{sepect_64}, middle), 
and the core-averaged field increases as $B\propto n^{2/3}_{\rm c}$ mainly due to the global compression (Fig. \ref{mag_evo} and Fig. \ref{cento_Eevo}c).
The field reaches equipartition on the Jeans scale at $n_{\rm c}\sim 10^{11}\ \rm cm^{-3}$ (Fig. \ref{sepect_64}, middle).
Thereafter, the field structure becomes coherent (as seen in Fig. \ref{turb_snap})
and the field energy exceeds the turbulent energy on the Jeans scale (middle panel of Fig. \ref{sepect_64}), with the turbulent motion damped by the coherent field.
Similarly, the core-averaged magnetic energy exceeds the core-averaged turbulent energy after reaching equipartition (Fig. \ref{cento_Eevo}c).

Finally, in the sub-Alfv\'{e}nic case ($B_{\rm init}=10^{-5}\ \rm G$, Fig. \ref{cento_Eevo}d), 
the magnetic force quickly suppresses the turbulence, as the former has higher energy than the latter from the beginning.
The turbulence suppression is particularly pronounced on smaller scales (Fig. \ref{cento_Eevo30}d, blue).
After the disappearance of the turbulence, the field evolution is similar to the pure rotation cases (see Fig. \ref{rot_mag_evo} and Fig. \ref{mag_evo}).

We have seen that a strong magnetic field can be generated by the turbulent dynamo even from a weak initial field.
Whether the field affects the gas dynamics, however, depends on the field strength on a large scale.
For example, in the cases of $B_{\rm init}=10^{-7}$ and $10^{-6}\ \rm G$ (Fig. \ref{turb_snap}, second and third columns), we can see 
ordered structures created perpendicular to the coherent field lines, suggesting that the gas inflows are significantly affected, particularly in later phases. 
By contrast, in the case of $B_{\rm init}=10^{-8}$ G  (Fig. \ref{turb_snap}, first column), where no coherent magnetic field is created during the collapse, the magnetic field has essentially no effect on the cloud structure.
These results suggest that a strong large-scale field needs to be generated in order to affect the gas dynamics.

Our results also show that launching MHD outflows needs a strong coherent field.
In the case of $B_{\rm init} = 10^{-6}\ \rm G$, 
outflows are launched by the coherent magnetic field in the pure rotation case  (Fig. \ref{no_turb_snap}, third column)
but not by the turbulent magnetic field (Fig. \ref{turb_snap}, third column),
even though the core-averaged magnetic field strength at the protostar formation is larger in the latter case than in the former.
In the case of $B_{\rm init}=10^{-5}\ \rm G$, 
the cloud collapse in the pure rotation and turbulent case proceeds in a very similar way,
as the magnetic field is so strong that its force quickly erases the turbulence in the latter case.
Therefore, outflows are launched by the strong coherent magnetic field even in the turbulent case in the same manner as in the pure rotation case. 
The condition for MHD outflow launching seen in our simulations
is in agreement with \cite{Gerrard2019}, who performed MHD simulations for the present-day star formation and found that a considerable initial uniform field component is required for the outflow launching.
\begin{figure}
\includegraphics[trim=0 60 40 80, scale = 0.5, clip]{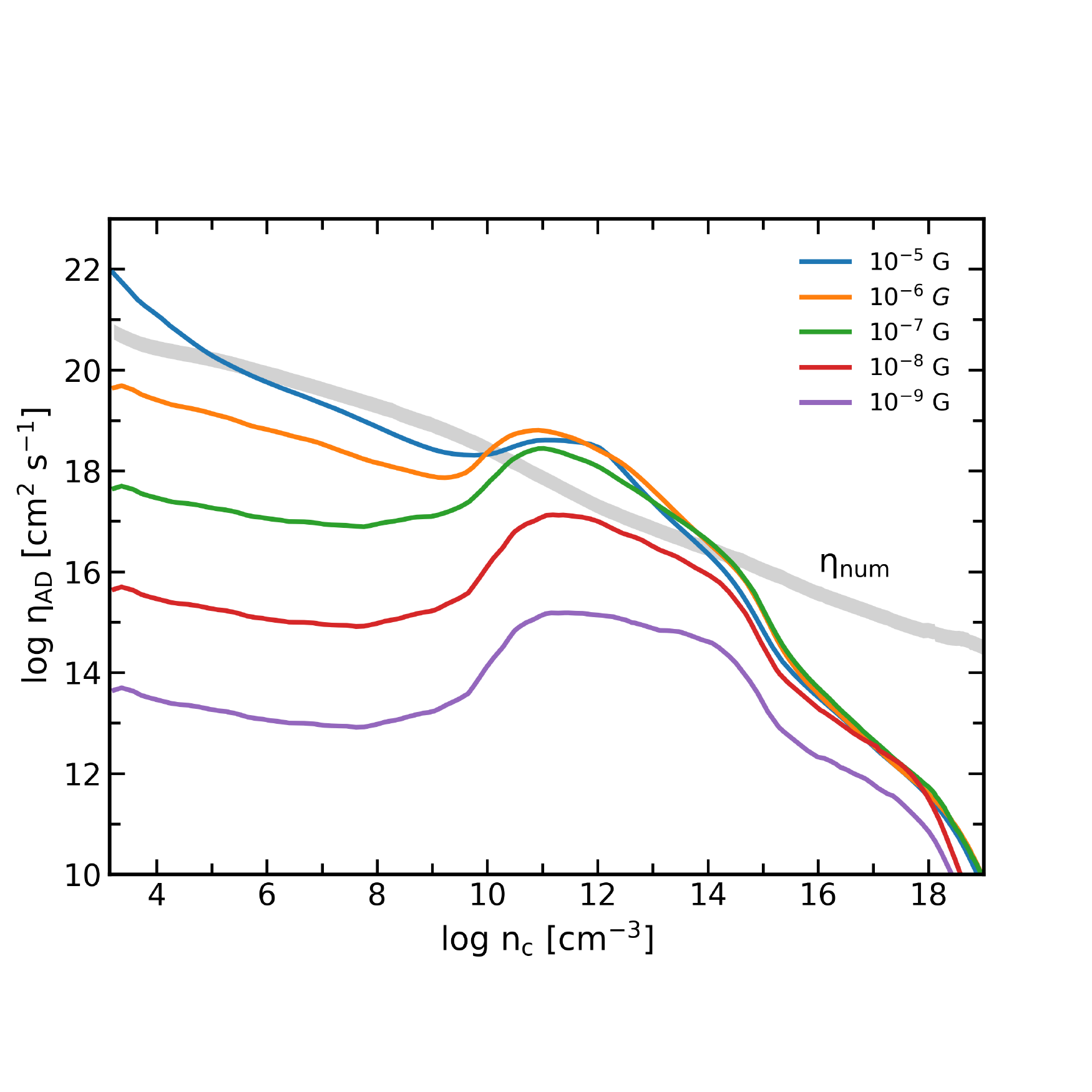}
\caption{The core-averaged AD resistivity as a function of the central density for the turbulent cases. The initial field strengths are indicated by the colors.
The gray shaded line represents the resistivity due to the finite numerical resolution.
}
\label{eta_evo}
\end{figure}
\begin{figure}
\centering
\includegraphics[trim=75 0 30 20, scale = 0.6, clip]{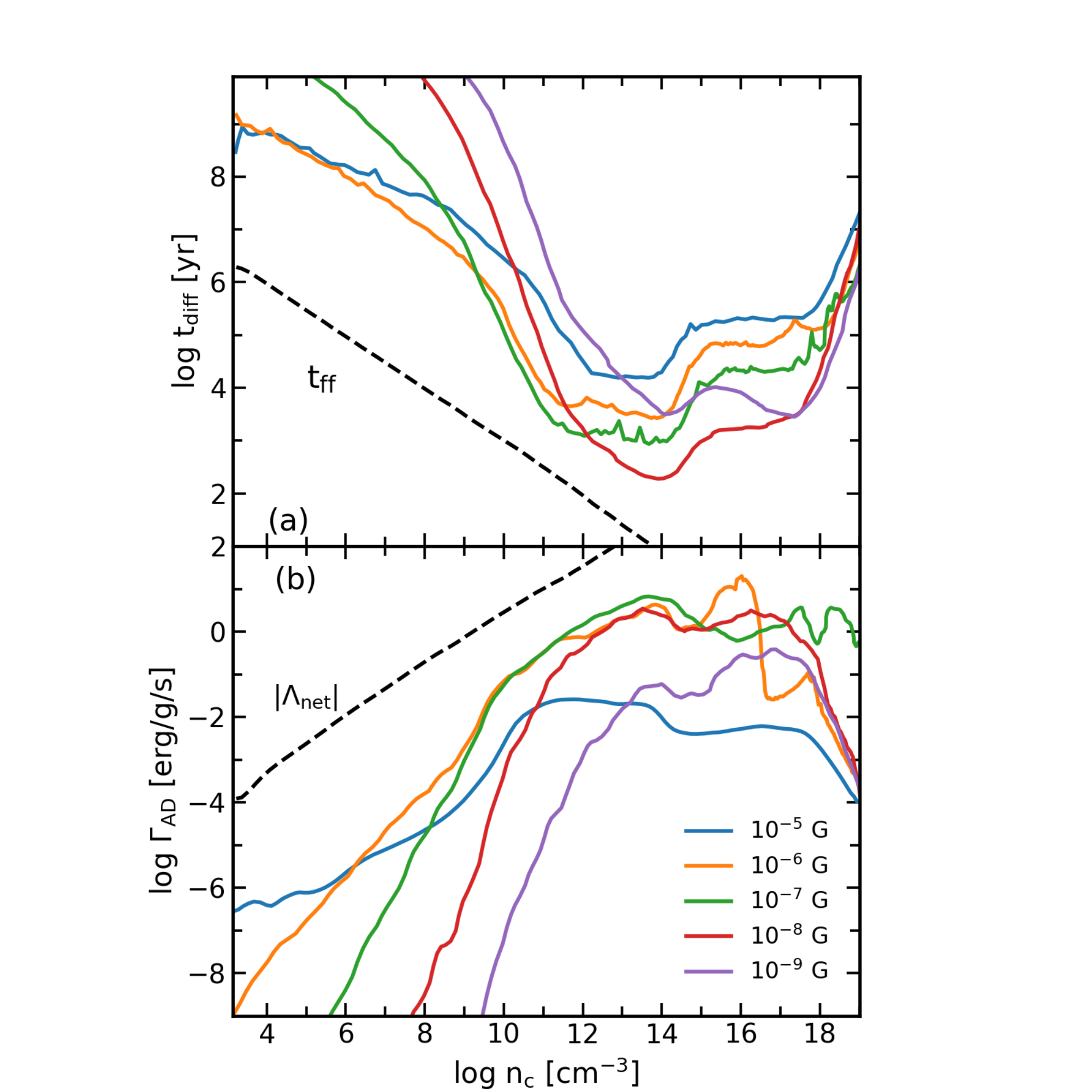} 
\caption{
(a) The comparison of the diffusion timescale (eq. \ref{eq_tdiff}, solid coloured) with the free-fall time (black dashed), and
(b) of the heating rate due to the ambipolar diffusion $\Gamma_{\rm AD}$ with the net cooling rate (black dashed), at the cloud centre.
The different colors indicate the different initial field strengths.}
\label{ADtime}
\end{figure}
\begin{figure*}
\includegraphics[trim=0 50 0 80, scale = 0.75, clip]{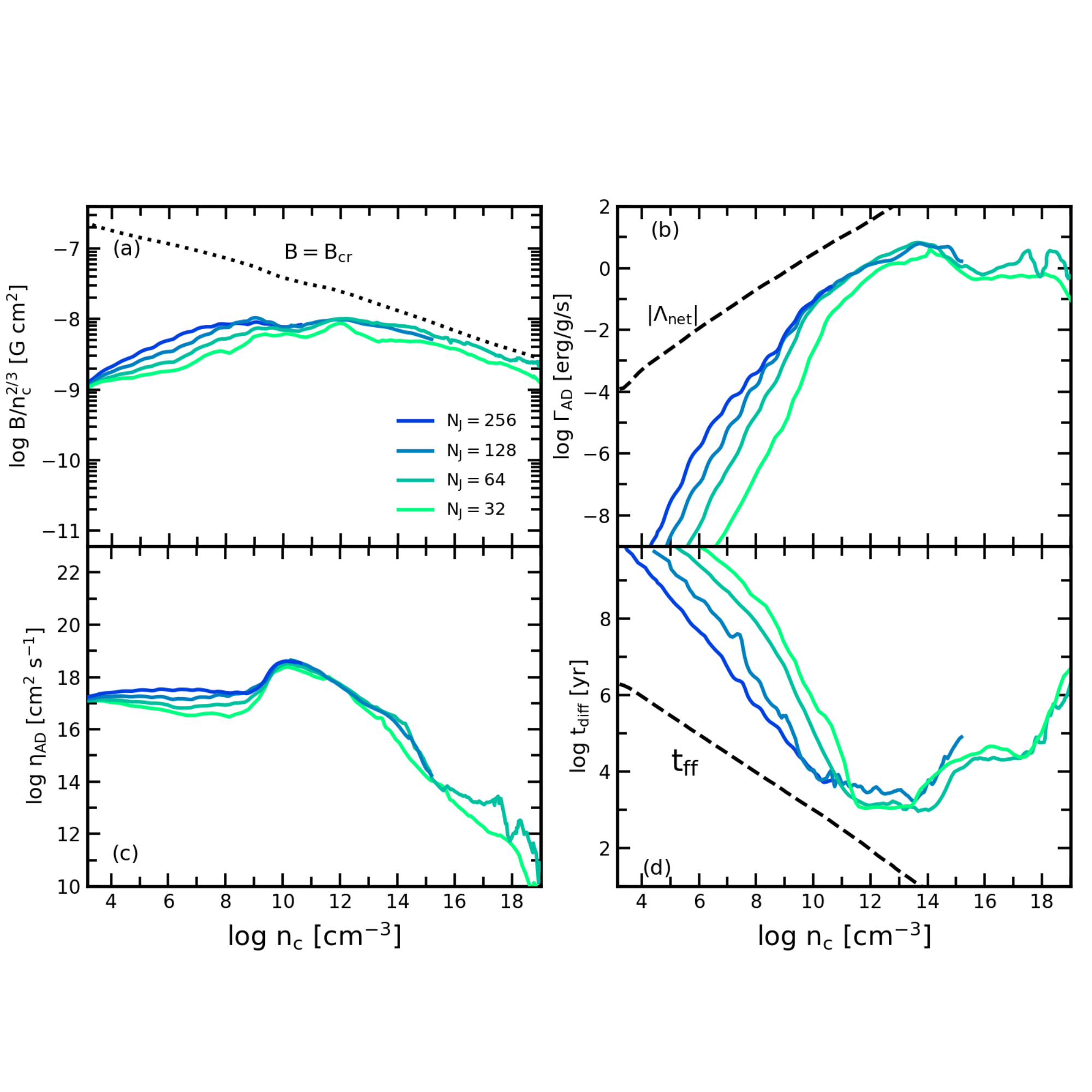} 
\caption{
The central evolution of (a) normalized magnetic field ($B/n_{\rm c}^{2/3}$), (b) resistivity of AD ($\eta_{\rm AD}$), (c) AD heating rate ($\Gamma_{\rm AD}$), and (d) diffusion timescale of AD ($t_{\rm diff}$) for the case of $B_{\rm init}=10^{-7}\ \rm G$ $(E_{\rm mag}/|E_{g}|=2\times10^{-5})$.
In each panel, we plot the different resolution cases of $N_{\rm J}=32,\ 64,\ 128,$ and $256$.
}
\label{evo_reso}
\end{figure*}
\subsection{Effect of ambipolar diffusion}
\label{sec_AD}
We have included ambipolar diffusion (AD) of magnetic field as a
dissipation mechanism in our simulations.  Here, to examine the AD
effects on the cloud evolution, we first see the evolution of the AD
resistivity with collapse, and then we compare simulations with and
without AD.

In this section, we will focus on the turbulent cases, in which the AD
effects tend to be stronger than in the pure rotation cases, as
explained below.  To begin with, the AD resistivity $\eta_{\rm AD}$
(eq.\ref{eq_etaAD}) can be approximated as
\begin{equation} \label{eq_AD}
\eta_{\rm AD} \simeq \frac{1}{4\pi \mu_{\rm i,n}}\left( \frac{B}{n_{\rm H}}\right)^2 \left( \langle \sigma v\rangle_{\rm i,n} y(\rm{n})y(\rm{e}) \right)^{-1},
\end{equation}
by considering only the contribution of the main charged species $\rm i$ and neutral species $\rm n$ (\citealp{Nakauchi2019}).
Then, keeping only the $B$ dependence of $\eta_{\rm AD}$ ($\propto B^2$), we can show
that the AD heating rate $\Gamma_{\rm AD}\ \rm[erg/g/s]$ has the dependence of $\propto B^{4}l^{-2}$,
with the field scale length $l$.  
As the AD heating is the process that transforms the magnetic energy to the thermal energy, the AD heating rate is related to the dissipation timescale:
\begin{equation} \label{eq_tdiff}
t_{\rm{diff}} = \frac{\mathcal{E}_{\rm mag}}{\Gamma_{\rm AD}} = \frac{B^2}{8 \pi \rho \Gamma_{\rm AD}},
\end{equation}
which has the dependence of $\propto B^{-2}l^{2}$.  These dependencies
imply that the AD heating, as well as the AD dissipation, is effective
if the magnetic field is strong on a small-scale where the dissipation
is active.  In the pure rotation cases, however, such a small-scale
field is not generated in the absence of the dynamo action, and thus the
AD effect is in general weaker than in the turbulent cases.  Actually, in
the pure rotation cases, we have confirmed that our results including
the AD effect are almost identical with those in the former ideal-MHD
simulations without the AD effect (\citealp{Sadanari2021}).

First, we see how the AD resistivity changes as the cloud collapse
proceeds.  Fig. \ref{eta_evo} shows the mass-weighted average of AD resistivity over the
central core as a function of the central density for the cases with
different initial field strengths.  We see that in the density range
$10^{10}-10^{14}\ \rm cm^{-3}$ the AD resistivity is enhanced because of
the low ionisation degree, as predicted by the one-zone calculations
(e.g, \citealp{Nakauchi2019}).  We also find a general trend that the AD
resistivity is higher with a stronger initial field, in line with the
relation $\eta_{\rm AD}\propto B^{2} $ (eq. \ref{eq_AD}).

Before going further, we need to keep in mind that MHD simulations are
always accompanied by artificial field dissipation on the smallest
scales due to the limited numerical resolution.  The AD resistivity
$\eta_{\rm AD}$ should be compared with this numerical resistivity
$\eta_{\rm num}$, which is $0.5-1$ times the kinematic viscosity $\nu_{\rm num}$ due to the limited numerical resolution (\citealp{Lesaffre_Balbus2007}).
Considering the smallest scale $l_{\rm\nu}$, where the turbulent velocity is given by $V_{\rm turb}(l_{\rm \nu})$,
the numerical kinematic viscosity can be estimated as
\begin{equation} \label{eq_nunum}
\nu_{\rm num} = l_{\rm \nu}\,V_{\rm turb}(l_{\rm \nu}),
\end{equation}
from the condition that the viscous dissipation timescale $t_{\rm
vis}\sim l^2_{\rm \nu}/\nu_{\rm num}$ equals to the eddy turnover time $t_{\rm
eddy}=l_{\rm \nu}/V_{\rm turb}(l_{\rm \nu})$.  As a representative
value, we plot the numerical resistivity $\eta_{\rm num}$ for the case
of $B_{\rm init}=10^{-7}\ \rm G$ with a thick grey line in
Fig. \ref{eta_evo}. Here, to evaluate $\nu_{\rm num}$ in
eq. (\ref{eq_nunum}), we substitute our simulation resolution $l_{\rm
\nu}\sim 1/(30k_{\rm J})$ and obtain $V_{\rm turb}(l_{\rm \nu})$ from
the Fourier analysis of the velocity field.
The numerical resistivity in the other cases is similar as long as the small-scale turbulence has a similar property. 
In the density range $10^{10}-10^{14}\ \rm cm^{-3}$, the AD resistivity is larger than the numerical resistivity except for the weak field cases with $B_{\rm init}\lesssim 10^{-8}\ \rm G$. 
Therefore, the AD effect in its active region ($10^{10}-10^{14}\ \rm cm^{-3}$), where the AD is most effective and potentially affects the evolution of cloud collapse, 
is well captured in our simulations, without significantly affected by the numerical resistivity.
We have also confirmed that the AD effect has no dependence on the resolution by performing simulations with different resolutions (see Sec. \ref{Appendix_A}).

Now, we compare simulations with and without the AD effect. AD
affects the cloud evolution in two ways: the magnetic field dissipation
and associated gas heating.  Some authors claimed that the thermal
evolution in the primordial gas during the collapse can be largely
affected by the AD heating based on one-zone calculations
(\citealp{Schleicher2009}; \citealp{Sethi2010}; \citealp{Nakauchi2019}).

Let us first see the AD dissipation effect on the growth of magnetic
field.  From Fig. \ref{cento_Eevo30}, we can see that the field energy
on the smallest scale of $k/k_{\rm J}=30$ in the cases with AD (red solid) tends to be slightly lower than in the cases without AD (red
dashed) in the AD active region of $10^{10}-10^{14}\ \rm cm^{-3}$.
This suggests that AD lowers the dynamo amplification on a small
scale although its effect is subtle.  In terms of the
core-averaged quantities, the effect is even weaker as can be seen from
Fig. \ref{cento_Eevo} (red solid and dashed). The inefficiency of the
AD dissipation can be understood by comparing the AD dissipation
timescale $t_{\rm{diff}}$ (eq. \ref{eq_tdiff}) with the free-fall time
$t_{\rm ff}$. The mass-weighted average of dissipation timescale $t_{\rm diff}$ over
the central region is plotted as a function of $n_{\rm c}$ in
Fig. \ref{ADtime} (a), along with the free-fall time $t_{\rm ff}$ (black
dashed).  We find that the free-fall time is always shorter than $t_{\rm
diff}$ for all the cases, i.e., AD is inefficient during the
collapse even in the AD active region where AD exceeds the numerical
dissipation ($10^{10}-10^{14}\ \rm cm^{-3}$).  Accordingly, the AD
effect does not significantly change the field back-reaction on the
turbulence, as we can see that the turbulence energy is hardly affected
either (Figs. \ref{cento_Eevo} and \ref{cento_Eevo30}; blue solid and
dashed).

Next, we see the effect of the AD heating on the thermal evolution.  In
Fig. \ref{cento_Eevo}, no difference is seen in the thermal energy (yellow)
between the cases with AD (solid) and without AD (dashed).  The reason
why the AD heating is so ineffective can be understood by comparing its
heating rate $\Gamma_{\rm AD}$ with the cooling rate.  Fig. \ref{ADtime}
(b) shows the mass-weighted average of AD heating rate $\Gamma_{\rm AD}$ (solid) over
the central region and the net cooling rate $|\Lambda_{\rm net}|$
(dashed).  We find that the AD heating rate is far lower than the
cooling rate at any density in all the cases, meaning that
the the AD heating plays only a negligible role in the thermal evolution.
We also find that the AD heating rate is always smaller than the
compressional heating rate (similar to the net cooling rate), partly
because of $t_{\rm diff}$ is longer than $t_{\rm ff}$, which reinforces
our claim that the AD heating is insignificant in thermal evolution.
Our results disagree with the
previous one-zone calculations (\citealp{Schleicher2009};
\citealp{Sethi2010}; \citealp{Nakauchi2019}) with respect to the effect of the AD heating. 
The difference can be attributed to stronger magnetic fields and collapse geometry assumed in their calculations than realized in our simulations.

In summary, the effects of both magnetic field dissipation and
associated gas heating due to AD are minor during the gravitational
collapse of the primordial gas.  Note, however, that this does not
exclude their possible importance in the later phase of first star
formation.
\subsection{Resolution dependence}\label{Appendix_A}
We check here whether our results presented above depend on the resolution. 
 Recall that the simulations were performed at a resolution of 64 Jeans length divisions (Jeans Number $N_{\rm J}=64$), which is high enough to capture the turbulent dynamo (\citealp{Federrath2011b}).
However, it is known that the results do not converge even with higher resolution (e.g., \citealp{Sur2010}; \citealp{Federrath2011b}; \citealp{Turk2012})
since the dynamo amplification rate is determined by the turbulent motion on the smallest scale.
Below, we show how much the magnetic field amplification, and hence the way AD works, varies with the resolution.
To this end, we performed the identical simulations but with different resolutions of $N_{\rm J}=32,\ 64,\ 128,$ and $256$ for the turbulent case with $B_{\rm init}=10^{-7}\ \rm G$.

Fig. \ref{evo_reso} shows the averaged central evolution of 
(a) normalized magnetic field ($B/n_{\rm c}^{2/3}$), (b) resistivity of AD ($\eta_{\rm AD}$), (c) AD heating rate ($\Gamma_{\rm AD}$), and (d) diffusion timescale of AD ($t_{\rm diff}$).
As we can see from the Fig. \ref{evo_reso}(a), 
the dynamo amplification during the kinematic stage is more efficient with higher resolution, as expected from the previous studies.
As a result, the transition timing from the kinematic to non-linear stage is earlier in higher resolution simulations.
For example, the transition takes place at $n_{\rm c}\sim 10^{7}\ \rm cm^{-3}$ in the case of $N_{\rm J}=256$, while it is delayed to $n_{\rm c}\sim 10^{9}\ \rm cm^{-3}$ in the case of $N_{\rm J}=64$.
After the transition, the magnetic field evolves as $B \propto n_{\rm c}^{2/3}$ regardless of resolution,
suggesting that the resolution dependence of non-linear amplification is weak.
Finally, after the non-linear stage, the field strength grows to the critical field value regardless of the resolution.

The resistivity of AD has a dependence of $\eta_{\rm AD} \propto B^{2} y(e)^{-1}$.
Since the thermal evolution does not change with the resolution, the resistivity $\eta_{\rm AD}$ varies only through the difference in the field strength,
resulting in higher $\eta_{\rm AD}$ with higher resolution during the kinematic stage  (Fig. \ref{evo_reso}c).  
After entering the non-linear stage, $\eta_{\rm AD}$ changes in the same manner in all the cases as the difference in the field strength becomes smaller.
In the density range where AD is the most active ($n_{\rm c}=10^{10}-10^{14}\ \rm cm^{-3}$), $\eta_{\rm AD}$ becomes almost the same irrespective of the resolution.
Consequently, the AD heating rate in Fig. \ref{evo_reso}(b) and diffusion timescale in Fig. \ref{evo_reso}(d) are also the same in this density range.
This suggests that the resolution dependence does not change
our conclusion that the AD has little effect either on the thermal evolution or the magnetic amplification during the collapse phase.
\section{Summary and Discussion}\label{sec4}
We have performed, non-ideal MHD simulations taking into account the ambipolar diffusion (AD) for the collapse of a first-star forming cloud core ($n_{\rm c}\sim 10^{3}\ \rm cm^{-3}$) up to the protostar formation ($n_{\rm c}\sim 10^{20}\ \rm cm^{-3}$).
We have studied the cases with different strengths of initial magnetic fields ($B_{\rm init}=10^{-9},\ 10^{-8},\ 10^{-7},\ 10^{-6},\ 10^{-5}\ \rm G$) and different initial velocity structure (pure rotation or rotation plus turbulence). 
Through the simulations, we have investigated how the magnetic field is amplified and affects the gas dynamics considering the AD effect.
Below, we summarize our findings.

\begin{itemize}
\item If a first-star forming cloud initially has purely rotational motion at a level expected from cosmological simulations, the cloud deforms to a sheet-like configuration regardless of the initial magnetic field strength, either by the centrifugal or magnetic force  (Fig. \ref{no_turb_snap}).
In a sheet-like cloud, the magnetic field is amplified by the gravitational compression at the same rate as the critical magnetic field $B_{\rm cr}$, 
defined as the field strength required for supporting the central core by the magnetic force (Fig. \ref{rot_mag_evo}).
Consequently, the initially weak field ($B_{\rm init}\ll B_{\rm cr,init}$) cannot catch up with $B_{\rm cr}$ during the collapse phase.
We have found that initially strong field ($B_{\rm init}\gtrsim 0.1\,B_{\rm cr,init}$, or $B_{\rm init} \geq 10^{-6}\ \rm G$ at $n_{\rm c}=10^{3}\ \rm cm^{-3}$) is needed
in order for the magnetic field to affect the gas dynamics either by magnetic braking or MHD outflows.
As the first-star forming regions are expected to have weak initial magnetic fields,
the magnetic field hardly affects the dynamics of a collapsing cloud in the pure rotation cases.

\item In reality, however, turbulence is naturally generated inside the cloud.
With the turbulence, the magnetic field is amplified not only by the gravitational compression, but also by the small-scale dynamo, resulting in a higher magnetic amplification rate than in the pure rotation cases  (Fig. \ref{mag_evo}).
Our simulations have shown that the magnetic field is amplified to the level of $B_{\rm cr}$ before protostar formation,
as long as the initial field has $B_{\rm init}\gtrsim 10^{-4}\,B_{\rm cr,init}$, or $B_{\rm init} \geq 10^{-9}\ \rm G$ at $n_{\rm c}=10^{3}\ \rm cm^{-3}$.

\item Our simulations suggest that it is a coherent magnetic field that has a significant impact on the cloud structure and velocity fields.  We find that a coherent field can be generated by the gravitational compression after the random magnetic fields reach equipartition up to the Jeans scale by the dynamo amplification.
Unless a coherent strong field is generated, MHD outflows such as magneto-centrifugal winds are not launched, at least immediately after protostar formation (Fig. \ref{turb_snap}).

\item In the case that a strong turbulent magnetic field is produced through the small-scale dynamo, AD dissipates the small-scale magnetic field in
the density range of $n_{\rm H}=10^{10-14}\ \rm cm^{-3}$, where the
resistivity is enhanced due to low ionisation degree.  However, the
field is dissipated only slightly because the dissipation timescale is
much longer than the free-fall timescale (Fig. \ref{ADtime}a).
Similarly, the AD heating does not affect the thermal evolution as the
AD heating rate is much smaller than the net cooling rate
(Fig. \ref{ADtime}b).  Our results contradict with previous one-zone
calculations that claimed that the AD heating significantly changes the
thermal evolution (\citealp{Schleicher2009}; \citealp{Sethi2010};
\citealp{Nakauchi2019}), but the discrepancy can be attributed to their
assumption of unrealistically strong magnetic field.  

\item Due to the inefficient AD, the field is amplified efficiently in primordial gas clouds unlike in the present-day case, where the dissipation is more efficient due to lower ionisation degree thanks to the presence of dust. As
a result, even though we assume a weak initial field for the primordial
gas cloud, the field near the protostar can reach $B=10^{3-5}\ \rm G$,
much stronger than in the present-day case.

\end{itemize}

Our simulations start from an idealistic initial condition of a Bonnor-Ebert sphere that mimics a first-star forming cloud core in the loitering phase (\citealp{Bromm1999}). 
On the other hand, 
Cosmological simulations (e.g., \citealp{Hirano2014}, \citeyear{Hirano2015}) suggest more complex density, temperature and velocity distributions and that their structure varies from one host minihalo to another.
\cite{Hirano2014} have shown that this structural diversity leads the difference in the collapse speed and the physical structure of gas envelope around protostars, thereby 
affecting the accretion history onto protostars.
Although the difference in collapse speed may cause some difference in the magnetic amplification rate, which needs further study in future, 
our conclusion that the AD does not affect magnetic field amplification would remain unchanged.

In the early  universe, 
we expect that weak seed fields  generated, e.g., by the Biermann battery mechanism (\citealp{Biermann1950}), grow exponentially by the small-scale dynamo in minihalos.
\cite{McKee2020} showed that growth timescale is sufficiently shorter than the virial timescale of a typical minihalo, 
implying that the magnetic field in the cloud core ($n_{\rm H}\sim 10^{3}\ \rm cm^{-3}$), similar to our initial condition, has already reached the equipartition level.
Judging from the results in the turbulent case of $B_{\rm init}=10^{-7}\ \rm G$, strong fields in the equipartition level are expected to be transformed into coherent fields by gravitational compression (Fig. \ref{sepect_64}), which would have a significant effect on gas accretion flows around protostars and the late accretion phase.

The accretion phase follows the protostar formation. 
Several authors have investigated the role of the magnetic field in the accretion phase
by performing 3D MHD simulations, but their results are seemingly
controversial so far. Simulations that follow the magnetic field amplification during the collapse showed that the
fields suppress disc fragmentation in the accretion phase, making IMF
top-heavy (\citealp{Sharda2020b}, \citeyear{Sharda2020a}; \citealp{Stacy2022}).
\citealp{Prole2022}, however, found no magnetic field
effects on the suppression of fragmentation in their simulations
starting from higher density but with equipartition random magnetic
fields with a small-scale dominant power spectrum, as predicted by the
dynamo theory for magnetic fields in the kinematic stage.  Their
difference may suggest that the magnetic field configuration in
the collapse phase significantly affects the later evolution in the
accretion phase. To reveal the magnetic field effect on the fragmentation, we
plan to extend our simulations to the accretion phase in a future work.

In our simulations, the forming protostar has a magnetic field of the
order of $10^{3}-10^{5}\ \rm G$, which is much stronger than the
expected value from MHD simulations of present-day star formation (e.g.,
\citealp{Machida2007}; \citealp{Vaytet2018}; \citealp{Machida_Basu2019};
\citealp{Wurster2022}).  Our simulations have shown that this is because
AD hardly dissipates the magnetic field in a primordial
collapsing cloud.  The generation of such a strong magnetic field around
the protostar can cause a variety of phenomena.  For example,
observations tell that many Class-0/I protostars in the present-day universe emit a large amount of energy ($10^{34-37}\ \rm erg$) in X-rays as protostellar
flares (e.g. \citealp{Tsuboi2000}; \citealp{Imanishi2001};
\citealp{Pillitteri2010}).  MHD simulations show that flares are generated when magnetic field energy stored on protostars by gas
accretion is released by magnetic reconnection (\citealp{Takasao2019}).
Other phenomena such as MRI-driven winds (\citealp{Suzuki_Inutsuka2014};
\citealp{Flock2011}; \citealp{Bai_Stone2013}) and coronal heating
(e.g. \citealp{washinoue2021}) may influence the protostellar evolution
and the temperature structure of the surrounding gas.  The consequences
of strong magnetic fields of primordial protostars will also be investigated
in a future work.

In the first star formation, the most studied mechanism that limits the
growth of protostars is ionizing feedback from protostars
(e.g. \citealp{McKee_Tan2008}; \citealp{Hosokawa2011}).  In the case of
present-day massive star formation, MHD outflows such as
magneto-centrifugal winds from a protostar or disc are also known to
play an important role in determining the final stellar mass (e.g.,
\citealp{Tanaka2017}; \citealp{Matsushita2017}; \citealp{Mignon2021}).
Our simulations have suggested that when turbulent magnetic fields
dominate, MHD outflows do not blow immediately after the protostar
formation, but may do so in the later accretion phase.  If the ionizing feedback reduces the gas density around the polar regions, the MHD outflows are no longer halted by the ram pressure of accreting
gas (e.g., \citealp{Machida2020}).  Oppositely, the MHD outflows may
remove the gas from the polar region and assist the expansion of ionized regions. In
any case, it is crucial to reveal the interplay between magnetic fields and
ionizing feedback. We will study this in the future by performing
radiation MHD simulations considering the ionizing feedback.
\label{Sec:discussion_conclusion}


\section*{Acknowledgments}
The authors wish to express their cordial gratitude to Prof. Takahiro Tanaka, the Leader of Innovative Area Grants-in-Aid for Scientific Research ``Gravitational wave physics and astronomy: Genesis'', for his continuous interest and encouragement.
The authors also would like to thank Drs. Gen Chiaki, Sunmyon Chon, Takashi Hosokawa, Ralf Klessen, Masahiro Machida, Hajime Susa, Masayuki Umemura and Naoki Yoshida for fruitful discussions and useful comments.
The numerical simulations were carried out on 
XC50  {\tt Aterui II} in Oshu City at the Center for Computational Astrophysics (CfCA) of the National Astronomical Observatory of Japan, 
the Cray XC40 at Yukawa Institute for Theoretical Physics in Kyoto University, and the computer cluster Draco at Frontier Research Institute for Interdisciplinary Sciences of Tohoku University.
This research is supported by Grants-in-Aid for Scientific Research (KO: 17H06360, 17H01102, 17H02869, 22H00149; KS: 21K20373; TM: 18H05437; KT: 16H05998, 21H04487) from the Japan Society for the Promotion of Science.  
KES acknowledges financial support from the Graduate Program on Physics for Universe of Tohoku University. 
KS appreciates the support by the Fellowship of the Japan Society for the Promotion of Science for Research Abroad and by the Hakubi Project Funding of Kyoto University.
KO acknowledges support from the Amaldi Research Center funded by the MIUR program "Dipartimento di Eccellenza" (CUP:B81I18001170001).
\section*{Data Availability}
The data underlying this article will be shared on reasonable request to the corresponding author.



\bibliography{hoge} 
\bibliographystyle{mnras}





\bsp	
\label{lastpage}
\end{document}